\documentclass[journal,10pt]{IEEEtran}

\usepackage{graphicx,amsmath,amssymb}
\usepackage{subfigure}
\usepackage{cite}
\usepackage{epstopdf}
\usepackage{fancyhdr}
\usepackage{mdwmath}
\usepackage{mdwtab}
\usepackage{balance}
\usepackage{xcolor}
\usepackage{bm}
\usepackage{amsthm}
\usepackage{algorithm}
\usepackage{multirow}
\usepackage{flafter}
\usepackage{setspace}
\usepackage{booktabs}
\usepackage{tabularx}
\usepackage{threeparttable}
\usepackage{stfloats}
\usepackage{algorithmic}
\usepackage{mathrsfs}
\usepackage{enumerate}
\usepackage{url}
\usepackage{caption}
\usepackage{lipsum}
\usepackage{gensymb}

\allowdisplaybreaks[4]

\newtheorem{lemma}{Lemma}

\newtheorem{corollary}{Corollary}

\newtheorem{assumption}{Assumption}

\newtheorem{guideline}{Guideline}

\newtheorem{definition}{Definition}

\newtheorem{remark}{Remark}

\newcommand{\tabincell}[2]{
    \begin{tabular}{@{}#1@{}}
        #2
    \end{tabular}
}

\begin{document}
\begin{sloppypar}

\title{DRL Enabled Coverage and Capacity Optimization in STAR-RIS-assisted Networks}

\author{{Xinyu~Gao,~\IEEEmembership{Member,~IEEE,}
Wenqiang~Yi,~\IEEEmembership{Member,~IEEE},\\
Yuanwei~Liu,~\IEEEmembership{Senior Member,~IEEE},
Jianhua~Zhang,~\IEEEmembership{Senior Member,~IEEE},\\
and Ping~Zhang,~\IEEEmembership{Fellow,~IEEE}

\thanks{Part of this work has been accepted to appear in the IEEE Wireless Communications and Networking Conference, Mar. 26–Mar. 29 March 2023~\cite{IEEEhowto:XGAO}.}
\thanks{X. Gao, W. Yi, and Y. Liu are with the Queen Mary University of London, London E1 4NS, U.K. (e-mail:\{x.gao,w.yi,yuanwei.liu\}@qmul.ac.uk).}}
\thanks{J. Zhang and P. Zhang are with the State Key Laboratory of Networking and Switching Technology, Beijing University of Posts and Telecommunications, Beijing 100876, China (email:\{jhzhang, pzhang\}@bupt.edu.cn).}

}
\maketitle

% \vspace{-1.7cm}
\begin{abstract}
  % \vspace{-0.3cm}
  Simultaneously transmitting and reflecting reconfigurable intelligent surfaces (STAR-RISs) is a promising passive device that contributes to full-space coverage via transmitting and reflecting the incident signal simultaneously. As a new paradigm in wireless communications, how to analyze the coverage and capacity performance of STAR-RISs becomes essential but challenging. To solve the coverage and capacity optimization (CCO) problem in STAR-RIS-assisted networks, a multi-objective proximal policy optimization (MO-PPO) algorithm is proposed to handle long-term effects. To strike a balance between each objective, the MO-PPO algorithm provides a set of optimal solutions to approach a Pareto front (PF), where the solution on the approximate PF is regarded as an optimal result. Moreover, in order to improve the performance of the MO-PPO algorithm, two update strategies, i.e., action-value-based update strategy (AVUS) and loss function-based update strategy (LFUS), are investigated. For the AVUS, the improved point is to integrate the action values of both coverage and capacity and then update the loss function. For the LFUS, the improved point is only to assign dynamic weights for both loss functions of coverage and capacity, while the weights are calculated by a min-norm solver at every update. The numerical results demonstrated that the investigated update strategies outperform the fixed weights MO optimization algorithms in different cases, which include a different number of sample grids, the number of STAR-RISs, the number of elements in the STAR-RISs, and the size of STAR-RISs. Additionally, the STAR-RIS-assisted networks achieve better performance than conventional wireless networks without STAR-RISs. Moreover, with the same bandwidth, a millimetre wave is able to provide higher capacity than sub-6 GHz, but at a cost of smaller coverage.
\end{abstract}

\vspace{-0.4cm}
\begin{IEEEkeywords}
  % \vspace{-0.3cm}
  Coverage and capacity optimization (CCO), multi-objective proximal policy optimization (MO-PPO), simultaneously transmitting and reflecting reconfigurable intelligent surfaces (STAR-RISs)
\end{IEEEkeywords}

\vspace{-0.3cm}
\section{Introduction}
For supporting increasing heterogeneous quality-of-service requirements of future wireless networks, e.g., high data rate, low latency, high reliability, massive connectivity, etc., an emerging communication paradigm, i.e., reconfigurable intelligent surfaces (RISs) \cite{IEEEhowto:YLiu1,IEEEhowto:JXu,IEEEhowto:XGAO1, IEEEhowto:EBasar} has been proposed to smartly control the wireless communication environment. RISs are able to offer line-of-sight (LOS) links to users located in blocked areas via reflection to improve both the coverage and capacity of conventional wireless networks. However, conventional RISs have maximal 180$\degree$ coverage, where the `blind zone' still exists at the backside of RISs. To overcome this limitation, a new concept named simultaneously transmitting and reflecting RISs (STAR-RISs) \cite{IEEEhowto:JXu1} becomes appealing. In contrast to conventional RISs, STAR-RISs are able to transmit and reflect the incident signal simultaneously, which contributes to full-space coverage \cite{IEEEhowto:CZhang}. As a new communication paradigm, it is an ultra-interesting question how STAR-RISs perform in terms of coverage and capacity. Note that coverage and capacity optimization (CCO) is one of the typical operational tasks mentioned by the 3rd Generation Partnership Project \cite{3GPP}. Since the coverage and capacity have several conflicting relationships, simultaneously optimizing them is important. For example, high transmit power contributes to large coverage but high inter-cell interference that reduces the capacity performance. To this end, multi-objective machine learning (MOML) \cite{IEEEhowto:EBalevi} algorithms can be a potential solution. Compared to single-objective algorithms, MOML algorithms are capable of handling the inherent conflict between objectives to achieve a group of optimal solutions by coordinating and compromising the requirements of objectives.

\vspace{-0.4cm}
\subsection{Related Works}
\vspace{-0.1cm}
\subsubsection{Capacity or Coverage Optimization for STAR-RISs Networks}
Conventional performance optimization for STAR-RIS-assisted networks focuses on a single objective: capacity or coverage. For capacity performance, there are some primary works. In \cite{Aldababsa2021}, a partitioning algorithm was proposed to determine the proper number of transmitting/reflecting elements that need to be assigned to each user and maximize the system sum-rate while guaranteeing the quality-of-service requirements for individual users. In STAR-RIS-assisted non-orthogonal multiple access (NOMA) systems, the authors in \cite{Zuo2021} proposed a sub-optimal two-layer iterative algorithm to maximize the achievable sum-rate by jointly optimizing the decoding order, power allocation coefficients, active beamforming, and transmission and reflection beamforming. The sum-rate performance of STAR-RIS-assisted full-duplex communication systems was investigated in \cite{Perera2022}, where the successive convex approximation technique has been employed to develop efficient algorithms for obtaining sub-optimal solutions. In \cite{Niu2022}, the authors proposed a sub-optimal block coordinate descent algorithm to maximize the weighted sum-rate for a STAR-RIS-assisted multiple-input multiple-output network. The authors in \cite{Wu2021} investigated the resource allocation problem in a STAR-RIS-assisted multi-carrier communication network and proposed location-based matching and semidefinite programming algorithms to maximize the system sum-rate. To derive the approximated average achievable rates of two users, the authors in \cite{Wang2021} investigated the performance of STAR-RIS-assisted downlink NOMA networks by a large array of analysis methods. For coverage performance, only one recent work has discussed its optimization problem. The STAR-RIS-assisted two-user communication networks were studied in \cite{Wu12021}, where the search-based algorithms were proposed to obtain the optimal one-dimensional (1D) coverage range.

% In \cite{Dreifuerst2021}, the authors developed and compared two approaches for maximizing coverage and minimizing interference by jointly optimizing the transmit power and downtilt (elevation tilt) settings across sectors.

\subsubsection{CCO based on MOML algorithms}
There are three main CCO solutions based on MOML algorithms: 1) Keep one objective in the objective function and move the rest objectives to constraints, while the obtained results are sub-optimal \cite{Dandanov2017}. 2) Assign a fixed weight to each objective. This method achieves the optimal results in a single scenario, while it cannot be used in other weight combinations, i.e., other network operation designs \cite{Skocaj2022}. 3) Obtain a set of optimal solutions according to Pareto-based multi-objective optimization algorithms, where one of these solutions can be selected to meet any specific network operation designs \cite{Dreifuerst2021}. More specifically, for the first method, an reinforcement learning (RL) algorithm-based solution for CCO by optimizing the base station (BS) antenna electrical tilt was proposed in \cite{Dandanov2017}, where the coverage objective was considered in the constraint. The proposed sub-optimal solution has the potential to reduce operational costs and complexity, as well as improve the quality of experience for mobile users. For the second method, in \cite{Skocaj2022}, minimization of drive tests (MDT)-driven deep RL algorithm was investigated to maximize the coverage and capacity by tuning antennas tilts on a cluster of cells from the cellular network, where the fixed weights were assigned for coverage and capacity. The results showed that the proposed MDT-driven approaches outperform baseline approaches, i.e., deep Q-network and best-first search, in terms of long-term reward and sample efficiency. For the third method, the authors in \cite{Dreifuerst2021} developed two RL algorithm-based approaches for maximizing coverage and minimizing interference by jointly optimizing the transmit power and antenna down-tilt across cells. The results suggested that data-driven techniques can effectively self-optimize coverage and capacity in cellular networks. There are some other promising MOML methods \cite{Yang2019, Sener2018}. A new algorithm was introduced in \cite{Yang2019} for multi-objective reinforcement learning (MORL) with linear preferences, with the goal of enabling few-shot adaptation to new tasks. The authors in \cite{Sener2018} proposed an upper bound for the multi-objective loss and show that it can be optimized efficiently. However, compared to a simple extension of the vanilla RL approaches to MOML algorithms, a new RL approach named proximal policy optimization (PPO) algorithm is able to provide a more stable training process (e.g., implement small batch updates in multiple training steps) and can be a booster for MOML algorithms.

% According to the current research, it can be obtained that it is usually necessary to train a decision policy to evaluate the performance of the wireless network, RL approaches (e.g., proximal policy optimization (PPO) algorithm) can be regarded as promising candidates. Additionally, there are some potential strategies to be integrated into RL algorithms. 

% The authors of \cite{Lin2019} proposed, validated, and productized a wireless ML-based scheme international organization for the standardization-based self-organizing network (SON) to mitigate cell coverage and interference problems in 4G Long Term Evolution (LTE) networks. 

\vspace{-0.4cm}
\subsection{Motivations and Contributions}
\vspace{-0.1cm}
As can be seen from related works, the CCO problem of STAR-RIS-assisted wireless networks is still in its early stage. In this research direction, there are two main challenges:
\begin{itemize}
  \item \textbf{Characterizing Coverage in STAR-RIS-assisted Networks}: STAR-RISs provide a new degree of freedom for manipulating signal propagation, thus increasing the flexibility of network design. Characterizing the two-dimensional (2D) coverage range for the STAR-RIS-assisted networks is challenging, compared to the one-dimensional coverage range described by the conventional networks. Additionally, the coverage characterization may be affected by the capacity, since the two objectives are conflicts.
  \item \textbf{Designing MORL Algorithms to Solve COO Problem}: Conventional Pareto-based MO optimization (MOO) solutions mainly aim to find an approximate Pareto front (PF) of objectives within a time step, which ignores the dynamic requirements of temporal correlations in long-term wireless communications. PPO is a policy gradient method where policy updates use a surrogate loss function to avoid catastrophic drops in performance. In addition, the new MOO methods \cite{Yang2019, Sener2018} have the capability to dynamically update the weights of objectives. Therefore, how to obtain the Pareto optimal (PO) solution based on the PPO algorithm and these two new MOO methods is challenging.
\end{itemize}
\par
To solve these challenges and fully reap the advantages of STAR-RISs, in this paper, we propose a new RL approach based on the PPO algorithm, named multi-objective PPO (MO-PPO) algorithm, to provide the maximum coverage and capacity for STAR-RIS-assisted networks. The optimal results obtained by the MO-PPO algorithm are different according to the different update strategies. The main contributions of this paper can be summarized as follows:
\begin{itemize}
  \item We propose a new model for a narrow-band downlink mode-splitting protocol-based STAR-RIS-assisted network consisting of two single-antenna BSs, where the serving range is defined as a square region. To quantitatively analyze the coverage and capacity, the serving range is discretized into numerous square grids, and the centre point of each grid sets as the evaluating sample point. Based on this framework, we formulate the CCO problem of STAR-RIS-assisted networks by jointly optimizing the transmit power, the reflection phase shift matrix, and the transmission phase shift matrix.
  \item We investigate an action value-based update strategy (AVUS) for the MO-PPO algorithm to solve the CCO problem. The core point of this strategy is to learn multiple policies for integrating the action values of both coverage and capacity by random sampling preferences, and further invoke a coefficient to update the policy by homotopy optimization. This update strategy with high performance is able to provide the optimal coverage and capacity, while it has to spend a long time to achieve convergence. Therefore, the AVUS has strict requirements on the computation resource, which is suitable for networks with strong computation capability.
  \item We adopt a loss function-based update strategy (LFUS) for the MO-PPO algorithm to reduce the complexity brought by the AVUS. The improved point is to assign dynamic weights for both loss functions of coverage and capacity and to update the whole MO-PPO policy with an integrated loss function of coverage and capacity. The dynamic weights are re-calculated by a min-norm solver at every update. Compared to the AVUS, this strategy has slightly worse performance, but it still has acceptable performance gain when compared to the conventional CCO solutions.
  \item We illustrate that both AVUS and LFUS-based MO-PPO algorithms are capable of striking a balance between the conflicting goals in terms of coverage and capacity. Then, AVUS and LFUS-based algorithms are able to provide the Pareto optimality compared to conventional fixed weights MOO algorithms. With the same bandwidth, a millimetre wave (mmWave) is able to provide better capacity while sub-6 GHz provides better coverage. Next, the coverage and capacity have a positive correlation with the number of STAR-RISs. Finally, when the number of elements in STAR-RISs is fixed, the coverage and capacity have a negative correlation with the physical size of STAR-RISs. 
\end{itemize}

\vspace{-0.7cm}
\subsection{Orgainization}
\vspace{-0.1cm}
The rest of this paper is organized as follows. Section II presents the system model for the considered STAR-RIS-assisted networks, and the coverage and capacity optimization problems are formulated. Section III provides the preliminaries, including the principles of the PPO algorithm and the PO solution. In Section IV, we investigate the two updated strategies-based MO-PPO algorithms, i.e., AVUS and LFUS, which are updated for different parts of the algorithm. Section V presents numerical results to verify the effectiveness of the proposed MO-PPO algorithms, by considering the different number of sample grids, the different number of elements in STAR-RISs, the different number of STAR-RISs, and the different physical sizes of STAR-RISs modules. Finally, Section VI concludes this paper.
\par
\emph{Notations:} Scalars, vectors, and matrices are denoted by lower-case, bold-face lower-case, and bold-face upper-case letters, respectively. The conjugate transpose of vector $\mathbf{a}$ is denoted by $\mathbf{a}^{H}$. The diag($\mathbf{a}$) denotes a diagonal matrix with the elements of vector $a$ on the main diagonal. The $||\mathbf{a}||$ denotes the norm of vector $\mathbf{a}$. The Mod($a,b$) denotes the modulus operation between values $a$ and $b$. The $\lfloor a \rfloor$ denotes the truncated argument of value $a$. The $*$ denotes the dot multiplication operation. The $\mathbb{E}[\mathbf{A}]$ is the expectation operator of matrix $\mathbf{A}$. The log$_{2}$($\mathbf{A}$) represents a logarithmic function with a constant base of 2 for matrix $\mathbf{A}$. The $\mathrm{tr}(\mathbf{A})$ denotes the trace of matrix $\mathbf{A}$.

\vspace{-0.5cm}
\section{System Model and Problem Formulation}
\vspace{-0.1cm}
\begin{figure*}[htbp]
  \setlength{\abovecaptionskip}{-0.1cm}
  \setlength{\belowcaptionskip}{-0.6cm}
  \centering
  \subfigure[]
  {
  \centering
  \includegraphics[scale = 0.3]{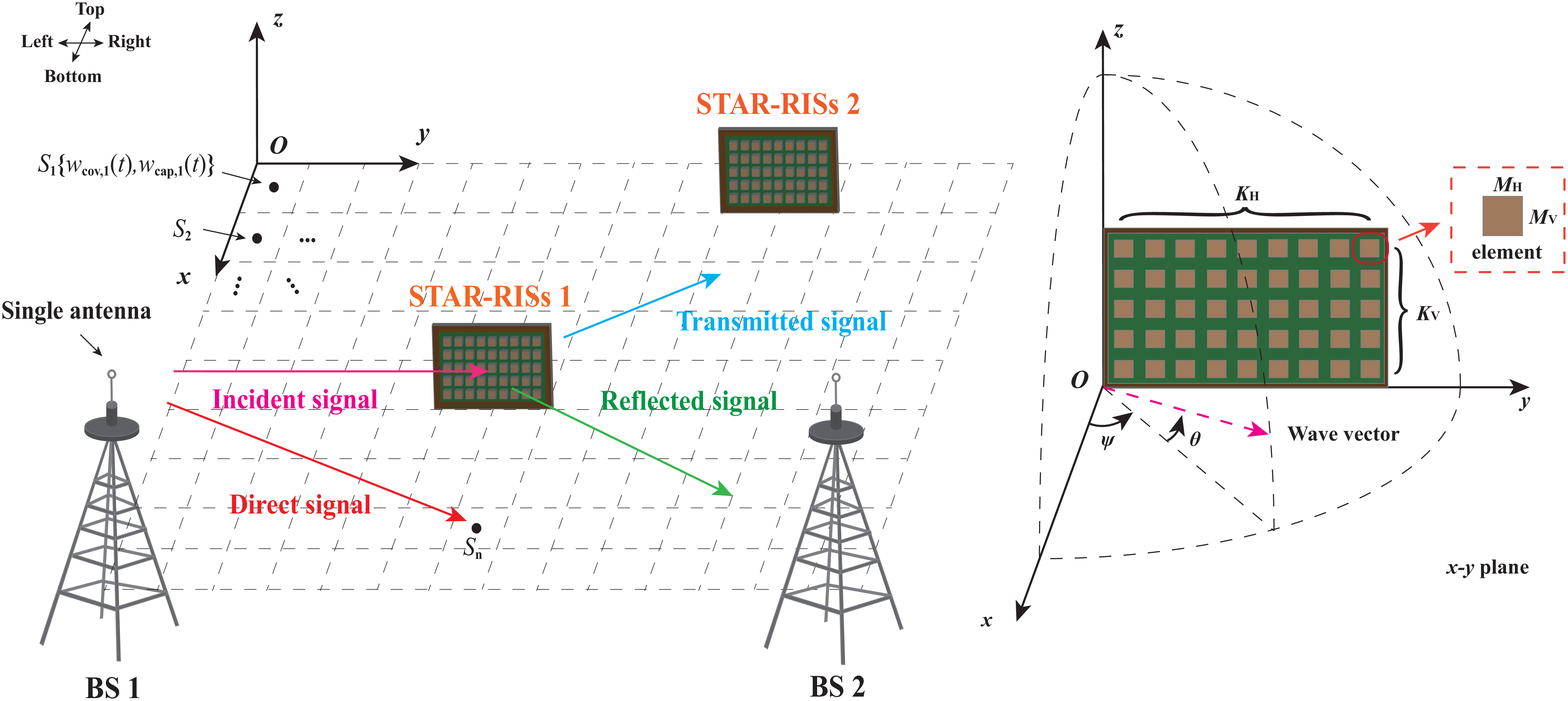}
  \label{system_model_all}
  }\\\vspace{-0.2cm}
  \subfigure[]
  {
  \centering
  \includegraphics[scale = 0.3]{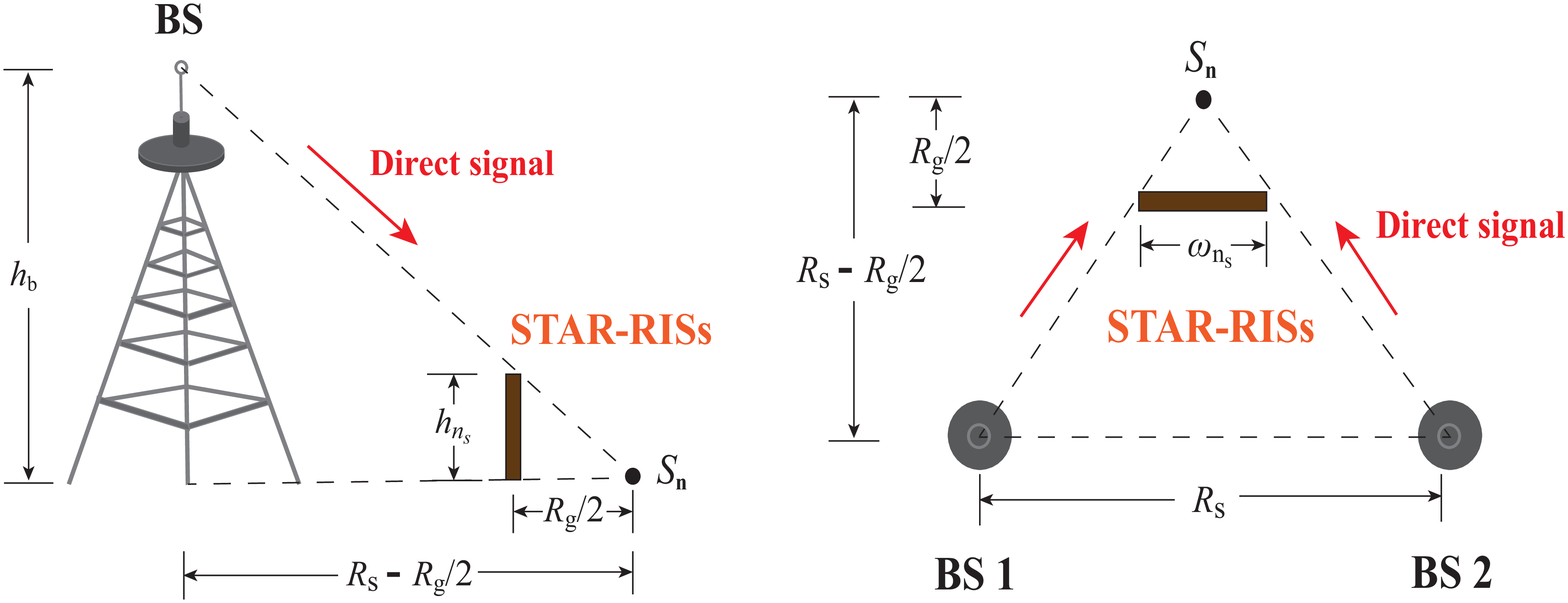}
  \label{system_model_RISsize}
  }
  \caption{Illustration of the considered narrow-band downlink STAR-RIS-assisted networks: (a) The geographic environment and the model of STAR-RISs; and (b) The constraints of height and width of STAR-RISs.}
  \label{system_model}
\end{figure*}

As shown in Fig.~\ref{system_model_all}, we consider a narrow-band downlink STAR-RIS-assisted network consisting of two single-antenna BSs and $N_s$ STAR-RISs of the same size equipped with $K = K_HK_V$ reconfigurable elements, where $K_H$ and $K_V$ denote the number of elements per row and column, respectively. The serving range is defined as a square region with the length of the side $R_s$, while the region is discretized into numerous square grids with the length of the side $R_g$, while the centre point of each grid acts as the sample point~\cite{0}. The BSs are located at the bottom left and bottom right corners with the same height $h_b$, while STAR-RISs with the height $h_{n_s}$ and width $\omega_{n_s}$ are deployed at designated locations in the square region. 

\vspace{-0.7cm}
\subsection{Grid-based Geographic Model}
\vspace{-0.1cm}
Assuming a three-dimensional (3D) Cartesian coordinate system, where the origin is set at the top-left corner. Here, the locations of two BSs and $n_s$-th STAR-RISs are denoted by $\mathrm{B}_1 = (R_s, 0, h_b)$, $\mathrm{B}_2 = (R_s, R_s, h_b)$, and $\mathrm{A}_{n_s} = (x_{n_s},y_{n_s},h_{n_s})$, respectively. Note that $h_{n_s}$ is the height of the STAR-RISs module, and the thickness of STAR-RISs is ignored. The height $h_{n_s}$ and width $\omega_{n_s}$ are depicted in Fig.~\ref{system_model_RISsize}. The indicators $\mathrm{I}_{h_{n_s}}$ and $\mathrm{I}_{\omega_{n_s}}$ are invoked to characterize $h_{n_s}$ and $\omega_{n_s}$ to further depict whether the direct links between the BSs and sample points exist or not, which can be expressed as follows:
\par
\vspace{-0.3cm}
\noindent
\begin{align}
  \mathrm{I}_{h_{n_s}} = \left\{ 
    \begin{array}{lr}\label{STAR-RISs height threshold}
      1, \hspace{1em} \mathbf{If} \hspace{0.5em} h_{n_s} \leq \frac{R_gh_b}{2R_s-R_g} \\ 
      0, \hspace{1em} \mathbf{If} \hspace{0.5em} h_{n_s} > \frac{R_gh_b}{2R_s-R_g}
    \end{array}
    \right., \\
  \mathrm{I}_{\omega_{n_s}} = \left\{ 
      \begin{array}{lr}\label{STAR-RISs width threshold}
        1, \hspace{1em} \mathbf{If} \hspace{0.5em} \omega_{n_s} \leq \frac{R_gR_s}{2R_s-R_g}\\ 
        0, \hspace{1em} \mathbf{If} \hspace{0.5em} \omega_{n_s} > \frac{R_gR_s}{2R_s-R_g}
      \end{array}
  \right..
\end{align}
\par
\vspace{-0.1cm}
\noindent
If we ensure that there is at least one direct link between BSs and any given sampling point, the indicators $\mathrm{I}_{h_{n_s}}$ and $\mathrm{I}_{\omega_{n_s}}$ need to satisfy the condition: $\mathrm{I}_{h_{n_s}} = 1$ and/or $\mathrm{I}_{\omega_{n_s}} = 1$. Thus, the indicators $\mathrm{I}_{h_{n_s}}$ and $\mathrm{I}_{\omega_{n_s}}$ can be unified as follows:
\par
\vspace{-0.1cm}
\noindent
\begin{align}\label{STAR-RISs size}
  \mathrm{I}_{n_s} = \mathrm{I}_{h_{n_s}} \lor \mathrm{I}_{\omega_{n_s}},
\end{align}
\par
\vspace{-0.1cm}
\noindent
where $\lor$ denotes the OR operator. $\mathrm{I}_{n_s} = 1$ denotes that the BSs are able to establish a direct link with the considered receiver from above and/or from the side of the STAR-RISs; Otherwise, there is no direct link between the BSs and the sample points. Additionally, if considering other numbers of BSs, this geographic model needs to be reconstructed according to the actual deployment of BSs.
\par
The coverage and capacity can be accordingly characterized based on the discretized grids. The total number of grids is $N = \lceil R_s/R_g \rceil^2$, where the set of sample points can be denoted as $\mathbf{s} = \{s_1,s_2,...,s_{N}\}$. In practical networks, in order to characterize the importance of each grid at each time step $t$, two time-related weights, $w_{\mathrm{cov}, i}(t)$ and $w_{\mathrm{cap}, i}(t)$, are assigned for coverage and capacity of each sample points $s_i$ ($i\in\{1, 2, \cdots, N\}$), respectively. Moreover, the weights have been unified, i.e., $\sum_{i=1}^{N}w_{\mathrm{cov}, s_i}(t) = 1$ and $\sum_{i=1}^{N}w_{\mathrm{cap}, s_i}(t) = 1$. In this system model, we study long-term communication with a time period $\mathcal{T}$. For each sample point at any time step, the weighted assignments $w_{\mathrm{cov}, s_i}(t)$ and $w_{\mathrm{cap}, s_i}(t)$ are influenced by the previous network performance and resource allocation strategy. Therefore, the considered problem can be regarded as a Markov Decision Process (MDP).

\vspace{-0.6cm}
\subsection{Spatially Correlated Channel Model}
\vspace{-0.1cm}
In this section, the fading channels from BSs to STAR-RISs, from STAR-RISs to sample points, and from BSs to sample points are introduced, as well as their spatial channel correlations. There are three different splitting protocols: 1) power-splitting (PS) protocol. In this case, all elements of the STAR-RIS are assumed to operate in the transmission and reflection modes, where the energy of the signal incident on each element is generally split into the energies of the transmitted and reflected signals with a ratio. 2) mode-splitting (MS) protocol. In this case, all elements of the STAR-RIS are divided into two groups. Specifically, one group contains some elements that operate in the transmission mode, while the other group contains the other elements operating in the reflection mode. 3) time-splitting (TS) protocols. Different from PS and MS, the TS STAR-RIS exploits the time domain and periodically switches all elements between the transmission mode and the reflection mode in different orthogonal time slots. In this paper, we consider the MS protocol in this paper\footnote{Compared to the PS and TS protocols for the STAR-RIS-assisted networks, MS is easier to implement \cite{IEEEhowto:JXu}.} \cite{IEEEhowto:YLiu1}. Denote $\mathbf{\Phi}_{\delta, n_s}$ as the coefficients of $n_s$-th STAR-RISs with mode $\delta$, where $\delta \in \{\mathrm{Re},\mathrm{Tr}\}$ represents the reflection and transmission modes. Due to the high path loss, this work assumes that signals are only reflected and transmitted by the STAR-RISs once. We consider the assumption of STAR-RISs with the same constant amplitude and continuous phase shifters in each mode\footnote{In practice, the transmission and reflection mode might be affected by circuit design based on different signal frequencies.}, where the phase shifters can be expressed as~\cite{2}: $\phi_{\delta,n_s,k_\delta} \in  [0, 2\pi), \forall k_\delta \in \{1,2,\cdots, K_\delta\}$, where $K_\delta$ is the total number of elements of STAR-RISs with mode $\delta$. The coefficients of $n_s$-th STAR-RISs are denoted as $\mathbf{\Phi}_{\delta, n_s} = [\mathbf{\Phi}_{\mathrm{Re}, n_s}, \mathbf{\Phi}_{\mathrm{Tr}, n_s}] = \mathrm{diag}(\sqrt{\beta_{\mathrm{Re},n_s}}e^{j\phi_{\mathrm{Re},n_s,1}}, ..., \sqrt{\beta_{\mathrm{Re},n_s}}e^{j\phi_{\mathrm{Re}, n_s,K_{\mathrm{Re}}}},  \sqrt{\beta_{\mathrm{Tr},n_s}} $\\$ e^{j\phi_{\mathrm{Tr},n_s,1}}, ..., \sqrt{\beta_{\mathrm{Tr},n_s}}e^{j\phi_{\mathrm{Re},n_s,K_{\mathrm{Tr}}}})$, where $\sqrt{\beta_{\delta,n_s}} \in (0, 1],\hspace{0.5em} K_{\mathrm{Re}} + K_{\mathrm{Tr}} = K,\hspace{0.5em} K_{\mathrm{Re}}\beta_{\mathrm{Re},n_s} + K_{\mathrm{Tr}}\beta_{\mathrm{Tr},n_s} = 1$~\cite{3}. As shown in Fig.~\ref{system_model_all}, a spherical coordinate system is defined with azimuth angel $\psi$ and elevation angel $\theta$ based on the 3D space. Denote the area of each element as $M = M_HM_V$, where $M_H$ and $M_V$ are the horizontal width and vertical height, respectively. The elements are deployed edge-to-edge in the isotropic scattering environment. Thus, the total area of $K$ elements can be expressed as $M_a = KM$. For the $k$-th element, its location can be expressed as \cite{r}:
\par
\vspace{-0.1cm}
\noindent
\begin{align}\label{4}
  \mathbf{l}_{k} = [0, x(k)M_H, y(k)M_V]^T,
\end{align}
\par
\vspace{-0.1cm}
\noindent
where $x(k)$ = mod($k-1, K_H$) and $y(k)$ = $\lfloor (k-1)/K_H \rfloor$ are the indices of $k$-th element. Assume a plane wave with wavelength $\lambda$ is impinging on the STAR-RISs, the array response vector is then given by:
\par
\vspace{-0.1cm}
\noindent
\begin{align}\label{5}
  \mathbf{a}(\psi, \theta) = [e^{j\mathbf{b}(\psi, \theta)^{T}l_{1}},e^{j\mathbf{b}(\psi, \theta)^{T}l_{2}},\cdots,e^{j\mathbf{b}(\psi, \theta)^{T}\mathbf{l}_{k}}]^T,
\end{align}
\par
\vspace{-0.1cm}
\noindent
where $\mathbf{b}(\psi, \theta) \in \mathbb{R}^{3 \times 1}$ is the wave vector, which can be expressed as follows:
\par
\vspace{-0.1cm}
\noindent
\begin{align}\label{6}
  \mathbf{b}(\psi, \theta) = \frac{2\pi}{\lambda}[\cos(\theta)\cos(\psi), \cos(\theta)\sin(\psi), \sin(\theta)]^T.
\end{align}
\par
\vspace{-0.1cm}
Assume that these channels are independently distributed and corresponding channel state information (CSI) is perfect. With this assumption, the obtained performance gains are regarded as the upper bound and every update of the phase shift of STAR-RISs is based on the instantaneous CSI. Denote $\mathbf{h}_{a,n_s}$, $\mathbf{h}_{\delta,n_s,s_i}$, and $h_{a,s_i}$ as the channel from $a$-th BS to $n_s$-th STAR-RISs with mode $\delta$, from $n_s$-th STAR-RISs to $s_i$-th sample point with mode $\delta$, and from $a$-th BS to $s_i$-th sample point, respectively. Here, the channels $h_{a,s_i}$ and $\mathbf{h}_{\delta,\mathrm{u}}, \mathrm{u} \in \{a,n_s;n_s,s_i\}$ can be modelled as Rician fading model, which is expressed as:
\par
\vspace{-0.1cm}
\noindent
\begin{align}
  h_{a,s_i} &= \sqrt{L_{a,s_i}} \Big( \sqrt{\frac{\alpha_{a,s_i}}{1+\alpha_{a,s_i}}}h_{a,s_i}^{\mathrm{LOS}} + \sqrt{\frac{1}{1+\alpha_{a,s_i}}}h_{a,s_i}^{\mathrm{NLOS}} \Big) ,\label{71} \\ 
  \mathbf{h}_{\delta,\mathrm{u}} &= \sqrt{L_{\mathrm{u}}} \Big( \sqrt{\frac{\alpha_{\mathrm{u}}}{1+\alpha_{\mathrm{u}}}}\mathbf{h}_{\delta,\mathrm{u}}^{\mathrm{LOS}} + \sqrt{\frac{1}{1+\alpha_{\mathrm{u}}}}\mathbf{h}_{\delta,\mathrm{u}}^{\mathrm{NLOS}} \Big), \label{73}
\end{align}
\par
\vspace{-0.1cm}
\noindent
where $L_{\overline{u}}, \overline{u} \in \{a,s_i;\mathrm{u}\}$, and $\alpha_{\overline{u}}, \overline{u} \in \{a,s_i;\mathrm{u}\}$ denote the corresponding path loss and Rician factor, respectively. The $h_{a,s_i}^{\mathrm{LOS}}$ denotes the deterministic LOS component of the channel from $a$-th BS to $s_i$-th sample point, which can be calculated according to the locations of BS and STAR-RISs. $\mathbf{h}_{\delta,a,n_s}^{\mathrm{LOS}} = \mathbf{b}(\psi^{\delta,a,n_s}, \theta^{\delta,a,n_s}) = \mathbf{b}\{\mathrm{arcsin}[ (h_{b}-h_{n_s}) /d_{a, n_s} ], \mathrm{arccos}[(R_s-x_{n_s})/\overline{d}_{\delta,a,n_s}]\}$ and $\mathbf{h}_{\delta,n_s,s_i}^{\mathrm{LOS}} = \mathbf{b}(\psi^{\delta,n_s,s_i}, \theta^{\delta,n_s,s_i}) = \mathbf{b}\{\mathrm{arcsin}(h_{n_s} /d_{\delta,n_s,s_i}), \mathrm{arccos}[(x_{n_s}-x_{s_i})/\overline{d}_{\delta,n_s,s_i}]\}$ are the deterministic LOS components for the channels from $a$-th BS to $n_s$-th STAR-RISs, and from $n_s$-th STAR-RISs to $s_i$-th sample point, respectively. Among them, $d_{\delta,a,n_s}$ and $d_{\delta,n_s,s_i}$ denote 3D distance between $a$-th BS and $n_s$-th STAR-RISs, and 3D distance between $n_s$-th STAR-RISs and $s_i$-th sample point, while $\overline{d}_{\delta,a,n_s}$ and $\overline{d}_{\delta,n_s,s_i}$ denote 2D distance between $a$-th BS and $n_s$-th STAR-RISs, and 2D distance between $n_s$-th STAR-RISs and $s_i$-th sample point. The $x_{n_s}$, $x_{s_i}$ indicate the $n_s$-th STAR-RISs, and $s_i$-th sample point, respectively. $\mathbf{h}_{\delta,a,n_s}^{\mathrm{NLOS}} \sim \mathcal{CN}\big(0, \mathbb{E}\big[\mathbf{h}_{\delta,a,n_s}^{\mathrm{NLOS}}(\mathbf{h}_{\delta,a,n_s}^{\mathrm{NLOS}})^{H}\big]\big)$, $\mathbf{h}_{\delta,n_s,s_i}^{\mathrm{NLOS}} \sim \mathcal{CN}\big(0, \mathbb{E}\big[\mathbf{h}_{\delta,n_s,s_i}^{\mathrm{NLOS}}(\mathbf{h}_{\delta,n_s,s_i}^{\mathrm{NLOS}})^{H}\big]\big)$, and $h_{a,s_i}^{\mathrm{NLOS}} \sim \mathcal{CN}(0, 1)$ are the non-line-of-sight (NLOS) components modeled as Rayleigh fading. Furthermore, for path loss $L_{\mathrm{u}}$, it can be modeled as $L_{\overline{u}} = C_0d_\mathrm{v}^{-\gamma_\mathrm{v}}, \mathrm{v} \in \{\{\delta,a,n_s\},\{\delta,n_s,s_i\},\{a,s_i\}\}$, where $C_0 = c/(4\pi d_0f_c)$ denotes the path loss at the reference distance $d_0 = 1$m under frequency $f_c$, $c$ is the velocity of light, and $\gamma_\mathrm{v}$ represents the path loss factor.

% $\mathrm{R}_{th}/(N w_{\mathrm{cov}, i}(t))$
\vspace{-0.4cm}
\subsection{Signal Model}
\vspace{-0.1cm}
Since the size of the STAR-RISs module affects the direct link, the received signal $y_{a,n_s,s_i} \in \mathbb{C}$ from the $a$-th BS to the $s_i$-th sample point via $n_s$-th STAR-RISs is determined by $\mathrm{I}_{n_{s}}$. Thus, the received signal $y_{a,n_s,s_i}$ can be written as \eqref{signal model}~\cite{5},
\par
\vspace{-0.2cm}
\noindent
\begin{figure*}[t]
\normalsize 
\begin{align}\label{signal model}
  y_{a,n_s,s_i} = \left\{ 
    \begin{array}{lr}
      \left(\mathbf{h}_{\delta,n_s,s_i}^\mathrm{H} \mathbf{\Phi}_{\delta, n_s} \mathbf{h}_{a,n_s} + h_{a,s_i}\right)x + n, \hspace{0.65em} \mathbf{If} \hspace{0.25em} \mathrm{I}_{n_{s}} = 1, \hspace{0.25em} \mathrm{I}_{\omega_{n_{s}}} = 0, \\ 
      \left(\mathbf{h}_{\delta,n_s,s_i}^\mathrm{H} \mathbf{\Phi}_{\delta, n_s} \mathbf{h}_{a,n_s} + \bar{a}_ah_{a,s_i}\right)x + n, \hspace{0.65em} \mathbf{If} \hspace{0.25em} \mathrm{I}_{n_{s}} = 1, \hspace{0.25em} \mathrm{I}_{h_{n_{s}}} = 0, \hspace{0.25em} \mathrm{I}_{\omega_{n_{s}}} = 1, \\
      \left(\mathbf{h}_{\delta,n_s,s_i}^\mathrm{H} \mathbf{\Phi}_{\delta, n_s} \mathbf{h}_{a,n_s}\right)x + n, \hspace{0.65em} \mathbf{If} \hspace{0.25em} \mathrm{I}_{n_{s}} = 0,
    \end{array}
    \right.
\end{align}
\hrulefill \vspace*{0pt}
\end{figure*}
\par
\vspace{-0.2cm}
\noindent
where the total transmit power $P_t = |x|^2$ and $n \sim \mathcal{CN}(0, \sigma^2)$ is the additive white Gaussian noise variance. $\bar{a}_a$ is a indicator that characterizing the direct link between $a$-th BS and $s_i$-th sample point. $\bar{a}_a = 1$ denotes that there is a direct link between $a$-th BS and $s_i$-th sample point, while $\bar{a}_a = 0$ denotes the direct link between $a$-th BS and $s_i$-th sample point is blocked. Due to the additional backhaul resources, the coordination of BSs needs extra communication requirements and computation resources, hence, we consider the most common practical strategy. based on the received signal power, the reference signal receiving power (RSRP) can be defined as the maximal useful signal power from all possible sources. The RSRP at the sample point $s_i$ is given by~\cite{0}:
\par
\vspace{-0.1cm}
\noindent
\begin{align}\label{8}
\mathrm{RSRP}_{s_i} = \max\limits_{a \in \{1, 2\}, n_s \in \{1, 2, \cdots, N_s\}} |y_{a,n_s,s_i} - n|^2.
\end{align}
\par
\vspace{-0.1cm}
\noindent
The achievable signal-to-interference-plus-noise ratio (SINR) of $s_i$-th sample point is calculated as follows~\cite{0}:
\par
\vspace{-0.1cm}
\noindent
\begin{align}\label{9}
  \mathrm{SINR}_{a,n_s,s_i}  = \frac{|y_{a,n_s,s_i} - n|^2}{\sum_{a^{'}=1,a^{'}\neq a}^{A}\sum_{n_s^{'}=1,n_s^{'}\neq n_s}^{N_s} |y_{a^{'},n_s^{'},s_i}-n|^2+{n^2}},
\end{align}
\par
\vspace{-0.1cm}
\noindent
where $a=1$, $a^{'}=2$; and $a=2$, $a^{'}=1$, otherwise. Assume the RSRP threshold for all sample points is $\mathrm{R}_{th}$, the weighted coverage ratio at time step $t$ can be written as
\par
\vspace{-0.1cm}
\noindent
\begin{align}\label{10}
  \mathrm{Coverage}(t) = \frac{||\mathbf{w}_{\mathrm{cov}, \check{\mathbf{s}}}(t) \cdot \check{\mathbf{s}}(t)||}{N},
\end{align}
\par
\vspace{-0.1cm}
\noindent
where $\check{\mathbf{s}}(t) = \{\check{s}_1(t),\check{s}_2(t),\cdots,\check{s}_{\tilde{N}}(t)\}$ is the set of the sample points at time $t$ that satisfying the condition $\mathrm{RSRP}_{\check{s}_{\tilde{n}}(t)} \geq \mathrm{R}_{th}, \check{s}_{\tilde{n}}(t) \in \check{\mathbf{s}}(t)$. $\mathbf{w}_{\mathrm{cov}, \check{\mathbf{s}}}(t) = \{w_{\mathrm{cov}, \check{s}_1}(t), w_{\mathrm{cov}, \check{s}_2}(t), \cdots, w_{\mathrm{cov}, \check{s}_{\tilde{N}}}(t)\}$ is the normalized corresponding coverage weights for the sample points $\check{\mathbf{s}}(t)$. For the network capacity, it is mainly determined by SINR, so at the time step $t$, the weighted capacity can be represented by
\par
\vspace{-0.1cm}
\noindent
\begin{align}\label{11}
\mathrm{Capacity}(t) & = \nonumber \\
& \sum_{s_i=1}^{N_s} w_{\mathrm{cap}, s_i}(t) \cdot \mathrm{B} \log_2\left(1+\mathrm{SINR}_{a^*, n_s^*,s_i}(t) \right),
\end{align}
\par
\vspace{-0.1cm}
\noindent
where $\mathrm{B}$ is the system bandwidth and $a^*, n_s^*= \arg\max_{a \in \{1, 2\}, n_s \in \{1, 2, \cdots, N_s\}} |y_{a,n_s,s_i} - n|^2$. According to equation \eqref{8}, when the transmit power $P_t(t)$ increases, the coverage will also increase with the increases of the RSRP. However, the interference in equation \eqref{9} will rise with the growth of the transmit power $P_t(t)$, while the performance of capacity will be degraded as the increase of interference. Therefore, there exists a conflict between coverage and capacity.

\vspace{-0.4cm}
\subsection{Problem Formulation}
\vspace{-0.1cm}
We focus on maximizing the long-term coverage and capacity for the whole serving area by optimizing the transmit power, the reflection phase shift matrix, and the transmission phase shift matrix. The formulated problem can be expressed as follows:
\par
\vspace{-0.1cm}
\noindent
\begin{align}
  \underset{P_t, \mathbf{\Phi}_{\mathrm{Re}, n_s},  \mathbf{\Phi}_{\mathrm{Tr}, n_s}}{\max} \hspace*{1em}& \int_{t=1} ^ \mathcal{T} \big[\mathrm{Coverage}(t), \mathrm{Capacity}(t)\big] \label{12}\\
  % {\rm s.t.} \hspace*{1em}& 0<N_s\le N_{s,\mathrm{max}}, \tag{\ref{12}{a}} \label{12a}\\
  \mathrm{s.\ t.} \hspace*{0.75em} & 0<P_t(t)\le P_{\mathrm{max}},\tag{\ref{12}{a}} \label{12a}\\
  % & \Phi_{n_s}\subset \mathbf{C}, \tag{\ref{12}{b}} \label{12b}\\
  & 0< \mathrm{tr}(\mathbf{\Phi}^H_{\delta,n_s}\mathbf{\Phi}_{\delta,n_s})< 1, \tag{\ref{12}{b}} \label{12b}\\
  & 0<\mathrm{tr}(\mathbf{\Phi}^H_{\mathrm{Re},n_s}\mathbf{\Phi}_{\mathrm{Re},n_s}) + \nonumber \\ 
  & \hspace{7em} \mathrm{tr}(\mathbf{\Phi}^H_{\mathrm{Tr},n_s}\mathbf{\Phi}_{\mathrm{Tr},n_s})\le 1, \tag{\ref{12}{c}} \label{12c}
  % & \sum\limits_{n_{s}=1}^{N_s}\Gamma_{n_{s}} \leq 1, \hspace{0.5em} \Gamma_{n_{s}} \in \Big\{0, \frac{1}{N_s} \Big\}, \tag{\ref{12}{e}} \label{12e}
\end{align}
\par
\vspace{-0.1cm}
\noindent
where $P_{\mathrm{max}}$ denotes the permitted maximum transmit power. Constraint \eqref{12a} limits the range of the transmit power. According to the energy conservation principle, constraints \eqref{12b} and \eqref{12c} show that both the energy of different modes and the sum energy of the reflected and transmitted signals is less than one. However, the main difficulty in solving the problem \eqref{12} owing to the following reasons. Firstly, the NLOS components for STAR-RIS-assisted links are hard to be determined before the STAR-RISs deployment. Secondly, the distribution weights $w_{\mathrm{cov}, s_i}(t)$ and $w_{\mathrm{cap}, s_i}(t)$ at time $t$ for calculating the coverage and capacity is not a continuous function. Thirdly, with respect to the continuous-time $t$, it's difficult to handle infinite variables optimization, since any adjacent time is subjected to the Markov chain. Thus, conventional non-convex optimization methods are not suitable for solving these difficulties. In the next section, the Pareto-based MO-PPO algorithms are invoked to solve this problem.

\vspace{-0.4cm}
\section{PO-based MO-PPO Algorithms}
\vspace{-0.1cm}
In this section, we first give a brief introduction to the principles of MDP definition, the PPO algorithm, and PO solution. Then the MDP in the MO-PPO algorithm is exhibited. Finally, the different update strategies of the PO-based MO-PPO algorithm are proposed to obtain the optimal policy applicable to the considered networks.

\vspace{-0.4cm}
\subsection{Basic Principles}
\vspace{-0.1cm}
\subsubsection{MDP Definition} For a typical RL problem, it can be defined as an MDP problem. A decision-maker called the agent will execute the action by interacting with the environment. The environment, in return, provides rewards and a new state based on the actions of the agent. In other words, RL presents the agent with rewards whether positive or negative based on its actions instead of teaching the agent how it should do something. Thus, the goal of RL is all about the goal to maximize the reward, where the MDP can be defined as the combination of the Markov reward process (MRP) with values judgement. Mathematically, we define MRP as:
\vspace{-0.1cm}
\begin{align}
  R_s = \mathbb{E}[R_{t+1}|S_t],
\end{align}
where the total reward $R_s$ OF MRP denotes the sum of reward $R_{t+1}$ gets from a particular state $S_t$. Then, the MDP can be defined as a tuple $\langle S, A, p, R\rangle$ with state space $S$, action space $A$, transition probability $p$, and reward $R$.

\subsubsection{PPO Algorithm}
\begin{figure*}[tbp]
  \setlength{\belowcaptionskip}{-0.6cm}
  \centering
  \includegraphics[scale = 0.35]{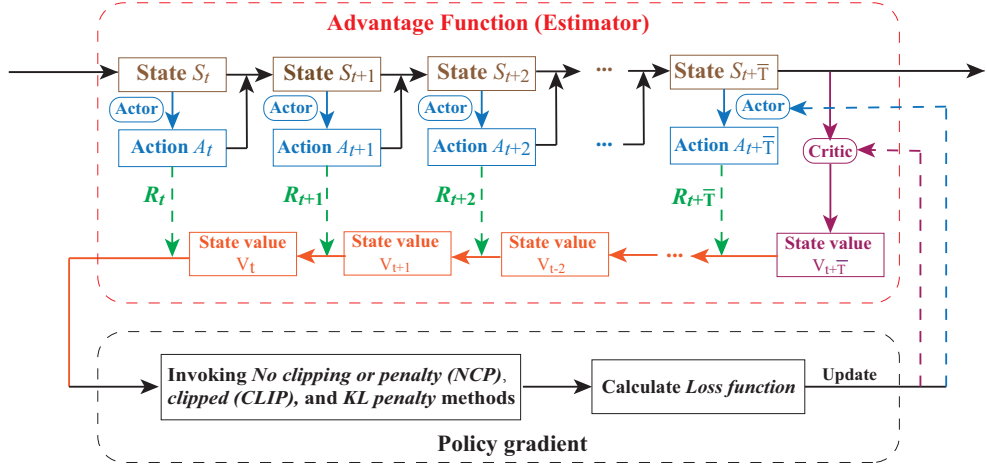}\\
  \caption{The framework of PPO algorithm.}\label{PPO}
\end{figure*}
PPO algorithm is based on trust region policy optimization \cite{TPRO} and utilizes the typical actor-critic architecture. The actor network is to determine the action according to the current state and the critic network, whereas the critic network is to evaluate how well the actor network performs the action. The configuration of the PPO algorithm is shown in Fig.~\ref{PPO}. Note that, the design of the architecture is modularized to separate the cohesion between neural networks, PPO algorithm, and environments. The action-value function in the PPO algorithm is replaced by an advantage function $\hat{A}_t$ at every $\overline{T}$ time steps, which is expressed as: 
\vspace{-0.1cm}
\begin{align}\label{PG}
  \hat{A}_t = \sum_{t=1}^{\overline{T}} \mathbf{Q}_{\pi_{\overline{\theta}}}(\mathrm{S}_t, \mathrm{A}_t) - V_{\pi_{\overline{\theta}}}(\mathrm{S}_t),
\end{align}
\par
\vspace{-0.1cm}
\noindent
where $\mathbf{Q}_{\pi_{\overline{\theta}}}(\mathrm{S}_t, \mathrm{A}_t)$ is the action-value function at policy $\pi_{\overline{\theta}}$ with state $\mathrm{S}_t$ and $\mathrm{A}_t$. The $V_{\pi_{\overline{\theta}}}(\mathrm{S}_t)$ is state-value predicted by the critic network. The update solution of the loss function, i.e., No clipping or penalty (NCP), can be expressed as:
\par
\vspace{-0.1cm}
\noindent
\begin{align}\label{NCP method}
  \mathcal{L}^{\mathrm{NCP}} = \min_{\overline{\theta}}\mathbb{E}_{t}\Bigg[\mathrm{\frac{\pi_{\overline{\theta}^{*}}(S_t, A_t)}{\pi_{\overline{\theta}}(S_t, A_t)}\hat{A}_{t}}\Bigg],
\end{align}
\par
\vspace{-0.1cm}
\noindent
where $\pi_{\overline{\theta}^{*}}(\cdot)$ and $\pi_{\overline{\theta}}(\cdot)$ denote the current policy and old policy. Since a large difference between the new and old policies often leads to destructively large policy \cite{6}, there are other two methods invoked for the PPO algorithm, i.e., clipped (CLIP) and Kullback–Leibler (KL) penalty methods. The two methods can be directly expressed as:
\vspace{-0.1cm}
\begin{align}\label{CLIP method}
  &\mathcal{L}^{\mathrm{CLIP}} = \nonumber \\
  &\min_{\overline{\theta}}\mathbb{E}_{t}\Bigg[\mathrm{\frac{\pi_{\overline{\theta}^{*}}(S_t, A_t)}{\pi_{\overline{\theta}}(S_t, A_t)}\hat{A}_{t}, clip\Bigg(\frac{\pi_{\overline{\theta}^{*}}(S_t, A_t)}{\pi_{\overline{\theta}}(S_t, A_t)}, 1-\epsilon, \epsilon\Bigg)\hat{A}_{t}}\Bigg],
\end{align}
\vspace{-0.6cm}
\begin{align}\label{KL method}
  \mathcal{L}^{\mathrm{KL}} = \min_{\overline{\theta}} \mathbb{E}_t\Big[\frac{\pi_{\overline{\theta}^{*}}(S_t, A_t)}{\pi_{\overline{\theta}}(S_t, A_t)}\hat{A}_t - \tilde{\beta}\mathrm{KL}(\pi_{\overline{\theta}^{*}}(S_t), \pi_{\overline{\theta}}(S_t))\Big],
\end{align}
\par
\vspace{-0.1cm}
\noindent
where $\epsilon$ is the probability ratio for the clipped method, and $\tilde{\beta}$ is a adjustment penalty coefficient for KL method.

\begin{figure}[tbp]
  \setlength{\belowcaptionskip}{-0.6cm}
  \centering
  \includegraphics[scale = 0.23]{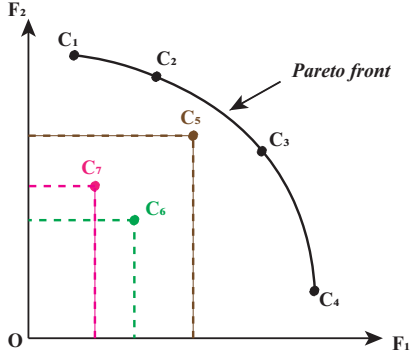}
  \caption{The Pareto solutions for two objectives.}\label{Pareto_front}
\end{figure}

\subsubsection{PO Solution}
In multi-objective optimization problems, each objective function may have an individual optimal solution, while these solutions usually have significant differences. Therefore, multi-objective optimization with such conflicting objective functions provides a set of optimal solutions, namely, PO solutions \cite{Pareto}. As shown in Fig.~\ref{Pareto_front}, considering two conflict objectives, both of which aim to be maximized. The point $\mathrm{C}_1$ represents a solution that $\mathrm{F}_2$ is near-maximum, but $\mathrm{F}_1$ is low, while point $\mathrm{C}_4$ indicates a solution $\mathrm{F}_1$ is near-maximum, but $\mathrm{F}_2$ is small. However, it is difficult to distinguish whether solution $\mathrm{C}_1$ is better than $\mathrm{C}_4$, or vice versa. In fact, there exist many such solutions belonging to the PO set, which forms a primary PF. Additionally, $\mathrm{C}_5$, $\mathrm{C}_6$, and $\mathrm{C}_7$ are the feasible solutions. $\mathrm{C}_5$ belongs to the second PF, while $\mathrm{C}_6$ and $\mathrm{C}_7$ are the part of the third PF \cite{Pareto}.

\vspace{-0.4cm}
\subsection{MO-PPO Framework}
\vspace{-0.1cm}
In this work, the locations of STAR-RISs are randomly pre-selected, and the locations of STAR-RISs are not overlapped. The locations stay the same before the training process achieves convergence. Moreover, STAR-RISs are placed along the y-axis direction\footnote{If the STAR-RISs are rotated, the geographic model and system model need to be reconstructed.} to ensure that transmit signal from any BS is reflected and transmitted using the same planar of each STAR-RISs.
\par
In the MO-PPO algorithm, the MDP can be represented by a tuple $\langle \overline{\mathbf{S}}, \overline{\mathbf{A}}, \mathbf{p}, \overline{\mathbf{R}}\rangle$ with state space $\overline{\mathbf{S}}$, action space $\overline{\mathbf{A}}$, reward space $\overline{\mathbf{R}}$. The $\mathbf{p}$ is the transition probability matrix indicating the probability of changing the current state to the next state. Define a controller as an agent, which controls both two BSs, to develop the policy from the BSs to sample points via STAR-RISs, i.e., the adjustment policies of phase shifts and transmit power. At each time step $t$, the controller observes the state $\mathbf{S}_t$ from state space $\overline{\mathbf{S}}$, and carries out an action $\mathbf{A}_t$ from action space $\overline{\mathbf{A}}$. The received reward $\mathbf{R} \subseteq \overline{\mathbf{R}}$ is calculated by the current state and action and determines the transition probability to the next state $\mathbf{S}_{t+1}$. Additionally, since the locations of STAR-RISs are pre-determined, the distance between any BS and $n_s$-th STAR-RISs is fixed. The coverage and capacity are determined by the distance between STAR-RISs and $s_i$-th point and the corresponding phase shift of the STAR-RISs, according to the \eqref{10} and \eqref{11}. Thus, the state $\mathbf{S}_{t}$ can be defined symbolically as follows:
\par
\vspace{-0.1cm}
\noindent
\begin{align}\label{state space}
  \mathbf{S}_t = 
  \begin{bmatrix} 
    \beta_{\mathrm{Re},n_s}(t), \beta_{\mathrm{Tr},n_s}(t), \mathbf{\Phi}_{\mathrm{Re},n_s}(t), \mathbf{\Phi}_{\mathrm{Tr},n_s}(t), P_t(t)
  \end{bmatrix}.
\end{align}
\par
\vspace{-0.1cm}
\noindent
For the action $\mathbf{A}_t$, the $\beta_{\mathrm{Tr},n_s}$ of STAR-RISs is discretized with small step $z$ as numerous values between $(0, 1)$, while the $\beta_{\mathrm{Re},n_s}$ is determined by $(1-\beta_{\mathrm{Tr},n_s})$ based on the energy constraint policy mentioned in \cite{IEEEhowto:EBasar}. In the MO-PPO algorithm, the category distributions of available locations and phase shifts of STAR-RISs are constructed first. Then, the agent samples phase shifts as an action according to the probability determined by the actor network. The action $\mathbf{A}_t$ can be expressed as follows:
\par
\vspace{-0.1cm}
\noindent
\begin{align}\label{action space}
  \mathbf{A}_t = 
  \begin{bmatrix} 
    \Delta \beta_{\mathrm{Re},n_s}, \Delta \beta_{\mathrm{Tr},n_s}, \Delta \phi_{\mathrm{Re},n_s}, \Delta \phi_{\mathrm{Tr},n_s}, \Delta P_t
  \end{bmatrix}.
\end{align}
\par
\vspace{-0.1cm}
\noindent
where $\Delta \beta_{\mathrm{Re},n_s} \in \{z,2z,\cdots,1-z\}$, $\Delta \beta_{\mathrm{Tr},n_s} \in \{1-z,1-2z,\cdots,z\}$, and $\Delta \phi_{\delta,n_s} \in \{ \phi_{\delta,n_s,1},\phi_{\delta,n_s,2},\cdots,\phi_{\delta,n_s,K_\delta}\}$ denote the possible values for the transmission amplitude, reflection amplitude, and possible phases for $n_s$-th STAR-RISs with mode $\delta$, respectively. The $\Delta P$ is chosen from $[0, zP_{\mathrm{max}}, 2zP_{\mathrm{max}},\cdots,P_{\mathrm{max}}]$. For $k$-th element, the phase is randomly selected from $[0,2\pi)$. To obtain the maximum transmission coverage and capacity that BSs is able to achieve in the time period $\mathcal{T}$, the reward is denoted as the difference of coverage $\Delta \mathrm{Cov}_{t \rightarrow t+1}$ and capacity $\Delta \mathrm{Cap}_{t \rightarrow t+1}$ in adjacent time steps, which can be calculated separately and expressed as a vector:
\par
\vspace{-0.1cm}
\noindent
\begin{align}\label{Multi-objective reward}
  \mathbf{R}_t(\mathbf{S}_t, \mathbf{A}_t) = 
  \begin{bmatrix}
    \Delta \mathrm{Cov}_{t \rightarrow t+1}, \Delta \mathrm{Cap}_{t \rightarrow t+1}
  \end{bmatrix}.
\end{align}
\par
\vspace{-0.1cm}
\noindent
Additionally, the loss function in the PPO algorithm can be evaluated according to \eqref{NCP method}, \eqref{CLIP method}, and \eqref{KL method}. In this work, we propose a novel framework for the MO-PPO algorithm, where two update strategies, i.e., AVUS and LFUS, are employed for the PO-based MO-PPO algorithms.

\vspace{-0.4cm}
\subsection{AVUS-based MO-PPO Algorithm}
\vspace{-0.1cm}
In this subsection, we consider the AVUS-based MO-PPO algorithm, where the MO-MDP can be rewritten as $\langle \overline{\mathbf{S}}, \overline{\mathbf{A}}, \mathbf{p}, \overline{\mathbf{R}}, \pmb{\Omega}, f_{\pmb{\Omega}}\rangle$. The $\pmb{\Omega}$ and $f_{\pmb{\Omega}}$ denote the preferences space and the functions of preference, respectively. In this case, a linear preference function is employed, i.e., $f_{\overline{\pmb{\omega}}}(\mathbf{R}_t(\mathbf{S}_t, \mathbf{A}_t)) = \overline{\pmb{\omega}}^T\mathbf{R}_t(\mathbf{S}_t, \mathbf{A}_t), \hspace{0.25em} \overline{\pmb{\omega}} \subseteq \pmb{\Omega}$. All possible returns from MO-MDP are able to form a PF $\mathcal{F}:=\{\hat{\mathbf{R}} \hspace{0.5em} | \hspace{0.5em} \forall \hspace{0.25em} \overline{\mathbf{R}} < \hat{\mathbf{R}}, \overline{\mathbf{R}} \subseteq \mathcal{F}^{*}\}$, where $\hat{\mathbf{R}}$ and $\overline{\mathbf{R}}$, and $\mathcal{F}^{*}$ denote the PO return, non-PO return, and the set of non-PO returns, respectively. For $\pmb{\Omega}$ in the AVUS, a PF-based convex coverage set $\mathbf{f}$ can be defined as:
\par
\vspace{-0.1cm}
\noindent
\begin{align}\label{PF-CCS}
  \mathbf{f} = \{\hat{\mathbf{R}} \subseteq \mathcal{F} \hspace{0.5em} | \hspace{0.5em} \exists \hspace{0.25em} \overline{\pmb{\omega}} \subseteq \pmb{\Omega}, \hspace{0.5em} \forall \hspace{0.25em} \overline{\mathbf{R}} \subseteq \mathcal{F}^{*}, \hspace{0.5em} \overline{\pmb{\omega}}^T\hat{\mathbf{R}} \geq \overline{\pmb{\omega}}^T\overline{\mathbf{R}} \}.
\end{align}
\par
\vspace{-0.1cm}
\noindent
The agent is able to learn a group of policies $\pmb{\Pi} = \{\pi_{\overline{\theta}^{1}}, \pi_{\overline{\theta}^{2}}, \cdots\}$ by interacting with the environments. Among them, there exists one linear preference vector $\overline{\pmb{\omega}}$ in policy $\pi_{\overline{\theta}^{*}}$ to satisfy:
\par
\vspace{-0.1cm}
\noindent
\begin{align}\label{learning phase}
  \overline{\pmb{\omega}}^T V^{\pi_{\overline{\theta}^{*}}}(\mathbf{S}_t) \geq \overline{\pmb{\omega}}^T V^{\pi_{\overline{\theta}}}(\mathbf{S}_t), \hspace{0.5em} \exists \hspace{0.25em} \overline{\pmb{\omega}} \subseteq \pmb{\Omega} ,
\end{align}
\par
\vspace{-0.1cm}
\noindent
where $V^{\pi_{\overline{\theta}^{*}}}(\mathbf{S}_t)$ denote the state-value function with state $\mathbf{S}_t$, and $\pi_{\overline{\theta}}$ denotes other any policy except $\pi_{\overline{\theta}^{*}}$. In the AVUS, the output network policy contains two sub-policies, which are optimized for coverage and capacity over different preferences, respectively. The core point of this strategy is to integrate the action values of all objectives, which are fully based on the convex envelope of the solution front. Here, we provide a theoretical analysis of the AVUS scheme below.

\subsubsection{Bellman Operator}
The standard single-objective PPO algorithm \cite{6} utilizes the Bellman expectation operator, where the action value function $Q_{\pi_{\overline{\theta}}}(S_t, A_t) $ by Bellman optimality operator $\mathcal{J}$ can be expressed as follows:
\par
\vspace{-0.1cm}
\noindent
\begin{align}\label{Bellman Operator}
  &(\mathcal{J}Q)_{\pi_{\overline{\theta}^{*}}}(\mathrm{S}_t, \mathrm{A}_t) = \nonumber \\
  &\hspace{3em} R_t(\mathrm{S}_t, \mathrm{A}_t)+\gamma\sum_{\mathrm{S}^{'} \subseteq \overline{\mathbf{S}}}p(\mathrm{S}_t, \mathrm{A}_t, \mathrm{S}^{'}) (\mathcal{H}Q)(\mathrm{S}^{'}, \mathrm{A}^{'}),
\end{align}
\par
\vspace{-0.1cm}
\noindent
where $\gamma$, $p(\mathrm{S}_t, \mathrm{A}_t, \mathrm{S}^{'})$, and $(\mathcal{H}Q)(\mathrm{S}^{'}, \mathrm{A}^{'}) = \max_{\mathrm{A}^{'} \subseteq \overline{\mathbf{A}}}Q_{\pi_{\overline{\theta}^{*}}}(\mathrm{S}^{'}, \mathrm{A}^{'})$ denote the discount factor, the transition probability from $\mathrm{S}_t$ to $\mathrm{S}^{'}$ by choosing $\mathrm{A}_t$, and the optimality filter for the next state $\mathrm{S}^{'}$, respectively. Then, we extend the single-objective PPO algorithm to the MO-PPO algorithm by considering an action-value function space $\pmb{\mathcal{Q}}$ to estimate expected total rewards under $\overline{\pmb{\omega}}$ preferences, where $\pmb{\mathcal{Q}}$ contains all bounded action-value functions $\mathbf{Q}(\mathbf{S}_t, \mathbf{A}_t, \overline{\pmb{\omega}})$. The corresponding value metric $\mathcal{D}$ can be defined as follows:
\par
\vspace{-0.1cm}
\noindent
\begin{align}\label{2D value space}
  &\mathcal{D}(\mathbf{Q}, \mathbf{Q}^{'}) \nonumber \\ 
  &= \max_{\mathbf{S}_t \subseteq \overline{\mathbf{S}}, \mathbf{A}_t \in \overline{\mathbf{A}}, \overline{\pmb{\omega}} \subseteq \pmb{\Omega}}||\overline{\pmb{\omega}}^T[\mathbf{Q}(\mathbf{S}_t, \mathbf{A}_t, \overline{\pmb{\omega}}) - \mathbf{Q}^{'}(\mathbf{S}_t, \mathbf{A}_t, \overline{\pmb{\omega}})]||,
\end{align}
\par
\vspace{-0.1cm}
\noindent
Based on any given policy $\pi_{\overline{\theta}}$, the evaluation operator of the action-value function in the MO-PPO algorithm can be defined as follows:
\par
\vspace{-0.1cm}
\noindent
\begin{align}\label{MO Operator}
  &(\mathcal{J}\mathbf{Q})_{\pi_{\overline{\theta}}}(\mathbf{S}_t, \mathbf{A}_t, \pmb{\Omega}) = \mathbf{R}_t(\mathbf{S}_t, \mathbf{A}_t) \nonumber \\ 
  &+\gamma\sum_{\mathbf{S}^{'} \subseteq \overline{\mathbf{S}}}p(\mathbf{S}_t, \mathbf{A}_t, \mathbf{S}^{'}) \sum_{\mathbf{A}^{'} \subseteq \overline{\mathbf{A}}} \pi_{\overline{\theta}}(\mathbf{A}^{'}|\mathbf{S}^{'}) \mathbf{Q}_{\pi_{\overline{\theta}}}(\mathbf{S}^{'}, \mathbf{A}^{'}, \overline{\pmb{\omega}}).
\end{align}
\par
\vspace{-0.1cm}
\noindent
Accordingly, we denote the optimality filter $\mathcal{H}$ for the MO-PPO action-value function as follows:
\par
\vspace{-0.1cm}
\noindent
\begin{align}\label{Filter}
  (\mathcal{H}\mathbf{Q})(\mathbf{S}^{'}, \mathbf{A}^{'}, \overline{\pmb{\omega}}) = \max_{A^{'} \subseteq \overline{\mathbf{A}}, \pmb{\Omega}^{'} \subseteq \overline{\pmb{\Omega}}} \pmb{\Omega}^T\mathbf{Q}_{\pi_{\overline{\theta}^{*}}}(\mathbf{S}^{'}, \mathbf{A}^{'}, \overline{\pmb{\omega}}^{'}).
\end{align}
\par
\vspace{-0.1cm}
\noindent
Intuitively, the filter $\mathcal{H}$ provides $\mathbf{Q}$ value under given $\mathbf{S}_t$ and $\overline{\pmb{\omega}}$ while handling the convex envelope of the solution front. The optimality operator $\mathcal{J}$ for the MO-PPO action function under optimal policy $\pi_{\overline{\theta}^{*}}$ can be defined as follows:
\par
\vspace{-0.1cm}
\noindent
\begin{align}\label{MO optimality filter-action}
  &(\mathcal{J}\mathbf{Q})_{\pi_{\overline{\theta}^{*}}}(\mathbf{S}_t, \mathbf{A}_t, \pmb{\Omega}) = \nonumber \\
  &\mathbf{R}_t(\mathbf{S}_t, \mathbf{A}_t)+\gamma\sum_{\mathbf{S}^{'}\subseteq \overline{\mathbf{S}}}p(\mathbf{S}_t, \mathbf{A}_t, \mathbf{S}^{'}) (\mathcal{H}\mathbf{Q})_{\pi_{\overline{\theta}^{*}}}(\mathbf{S}^{'}, \mathbf{A}^{'}, \overline{\pmb{\omega}}).
\end{align}
\par
\vspace{-0.1cm}
\noindent
Compared to \eqref{Bellman Operator}, \eqref{MO optimality filter-action} integrated all the objectives by invoking $\overline{\pmb{\omega}}$ to update the policy of each objective simultaneously.
\par
\vspace{-0.1cm}
\noindent
\begin{remark}\label{remark 1}
  Compared to the single objective optimization problem, the policy in AVUS contains a preference space $\overline{\pmb{\omega}}$, which is utilized to estimate the total rewards under multi-objective and update the whole policy. Note that, each objective has its own policy instead of sharing a common policy.
\end{remark}
\par
\vspace{-0.1cm}
\noindent
% \begin{remark}\label{remark 2}
%   The transition probability $p(\mathbf{S}_t, \mathbf{A}_t, \mathbf{S}^{'})$ is unknown since the environment is not known entirely, the loss function in the AVUS needs to be calculated with a different expression.
% \end{remark}

% \vspace{-0.3cm}
\subsubsection{Loss Function}
Typically the environment is not known entirely so there is no closed-form solution to obtain optimal action-value and state-value functions. In this case, the advantage estimator can be expressed as \eqref{origin advantage}, where $\mathbf{R}_{t}$ and $V_{\pi_{\overline{\theta}}}(.,.)$ denote the obtained reward at each time step, the output state-value by critic network, respectively. In our proposed strategy, the loss function can be calculated based on the NCP method, CLIP method, and KL penalty method, which are expressed as \eqref{Original Loss function1} - \eqref{Original Loss function3}.
\par
\vspace{-0.1cm}
\noindent
\begin{figure*}[htbp]
  \normalsize 
  \begin{align}\label{origin advantage}
    \hat{\mathbf{A}}_{t}^{\pi_{\overline{\theta}^{*}}}(\overline{\pmb{\omega}}) &= \sum_{t}^{\overline{T}}\mathbf{Q}_{\pi_{\overline{\theta}}}(\mathbf{S}_t, \mathbf{A}_t, \overline{\pmb{\omega}}) - V_{\pi_{\overline{\theta}}}(\mathbf{S}_t, \overline{\pmb{\omega}}) \nonumber \\
    &= \mathbf{R}_{t} + \gamma \mathbf{R}_{t+1} + \gamma^2 \mathbf{R}_{t+2} + \cdots + \gamma^{\overline{T}-t+1}\mathbf{R}_{\overline{T}-1} + \gamma^{\overline{T}-t}V_{\pi_{\overline{\theta}}}(\mathbf{S}^{\overline{T}}, \overline{\pmb{\omega}}) - V_{\pi_{\overline{\theta}}}(\mathbf{S}_t, \overline{\pmb{\omega}}),
  \end{align}
  \hrulefill \vspace*{0pt}
\end{figure*}
\begin{figure*}[htbp]
  \normalsize 
\begin{align}
  &\mathcal{L}^{\mathrm{NCP}}_{1}(\overline{\pmb{\theta}}, \overline{\pmb{\omega}}) = \mathbb{E}_{t}\Big\{\mathrm{min}\Big[\mathrm{\frac{\pi_{\overline{\pmb{\theta}}^{*}}(\mathbf{S}_t, \mathbf{A}_t, \overline{\pmb{\omega}})}{\pi_{\overline{\pmb{\theta}}}(\mathbf{S}_t, \mathbf{A}_t, \overline{\pmb{\omega}})}\hat{\mathbf{A}}_{t}^{\pi_{\overline{\pmb{\theta}}^{*}}}(\overline{\pmb{\omega}})}\Big]\Big\}, \label{Original Loss function1} \\
  &\mathcal{L}^{\mathrm{CLIP}}_{1}(\overline{\pmb{\theta}}, \overline{\pmb{\omega}}) = \mathbb{E}_{t}\Big\{\mathrm{min}\Big[\mathrm{\frac{\pi_{\overline{\pmb{\theta}}^{*}}(\mathbf{S}_t, \mathbf{A}_t, \overline{\pmb{\omega}})}{\pi_{\overline{\pmb{\theta}}}(\mathbf{S}_t, \mathbf{A}_t, \overline{\pmb{\omega}})}\hat{\mathbf{A}}_{t}^{\pi_{\overline{\pmb{\theta}}^{*}}}(\overline{\pmb{\omega}}),clip\Big(\frac{\pi_{\overline{\pmb{\theta}}^{*}}(\mathbf{S}_t, \mathbf{A}_t, \overline{\pmb{\omega}})}{\pi_{\overline{\pmb{\theta}}}(\mathbf{S}_t, \mathbf{A}_t, \overline{\pmb{\omega}})}, 1-\epsilon, \epsilon\Big)\hat{\mathbf{A}}_{t}^{\pi_{\overline{\pmb{\theta}}^{*}}}(\overline{\pmb{\omega}})}\Big]\Big\}, \label{Original Loss function2} \\
  &\mathcal{L}^{\mathrm{KL}}_{1}(\overline{\pmb{\theta}}, \overline{\pmb{\omega}}) = \mathbb{E}_{t}\Big\{\mathrm{min}\Big[\mathrm{\frac{\pi_{\overline{\pmb{\theta}}^{*}}(\mathbf{S}_t, \mathbf{A}_t, \overline{\pmb{\omega}})}{\pi_{\overline{\pmb{\theta}}}(\mathbf{S}_t, \mathbf{A}_t, \overline{\pmb{\omega}})}\hat{\mathbf{A}}_{t}^{\pi_{\overline{\pmb{\theta}}^{*}}}(\overline{\pmb{\omega}}),\tilde{\beta}KL(\pi_{\overline{\pmb{\theta}}^{*}}(\mathbf{S}_t, \overline{\pmb{\omega}}), \pi_{\overline{\pmb{\theta}}}(\mathbf{S}_t, \overline{\pmb{\omega}}))}\Big]\Big\}.\label{Original Loss function3}
\end{align}
\hrulefill \vspace*{0pt}
\end{figure*}
\par
\vspace{-0.3cm}
% \noindent
At each update, the optimal method will be selected as follows:
\par
\vspace{-0.1cm}
\noindent
\begin{align}\label{optimal0}
  \mathcal{L}^{\mathrm{optimal}}_{1}(\overline{\pmb{\theta}}, \overline{\pmb{\omega}}) = \max\{\mathcal{L}^{\mathrm{NCP}}_{1}(\overline{\pmb{\theta}}, \overline{\pmb{\omega}}), \mathcal{L}^{\mathrm{CLIP}}_{1}(\overline{\pmb{\theta}}, \overline{\pmb{\omega}}), \mathcal{L}^{\mathrm{KL}}_{1}(\overline{\pmb{\theta}}, \overline{\pmb{\omega}})\}.
\end{align}
\par
\vspace{-0.1cm}
\noindent
However, owing to a large number of discrete solutions in the optimal PO front, directly optimizing $\mathcal{L}^{NCP/CLIP/KL}_{1}(\overline{\pmb{\theta}}, \overline{\pmb{\omega}})$ in practice is still challenging. To address the difficulty, auxiliary loss functions are invoked and the optimal selection is expressed as follows:
\par
\vspace{-0.1cm}
\noindent
\begin{align}\label{weight Loss function}
  &\mathcal{L}^{\mathrm{Optimal}}_{2}(\overline{\pmb{\theta}}, \overline{\pmb{\omega}}) \nonumber \\
  &=\max\{\mathcal{L}^{\mathrm{NCP}}_{2}(\overline{\pmb{\theta}}, \overline{\pmb{\omega}}), \mathcal{L}^{\mathrm{CLIP}}_{2}(\overline{\pmb{\theta}}, \overline{\pmb{\omega}}), \mathcal{L}^{\mathrm{KL}}_{2}(\overline{\pmb{\theta}}, \overline{\pmb{\omega}})\}, \nonumber \\
  &= \max\{\overline{\pmb{\omega}}^T\mathcal{L}^{\mathrm{NCP}}_{1}(\overline{\pmb{\theta}}, \overline{\pmb{\omega}}), \overline{\pmb{\omega}}^T\mathcal{L}^{\mathrm{CLIP}}_{1}(\overline{\pmb{\theta}}, \overline{\pmb{\omega}}), \overline{\pmb{\omega}}^T\mathcal{L}^{\mathrm{KL}}_{1}(\overline{\pmb{\theta}}, \overline{\pmb{\omega}})\}.
\end{align}
\vspace{-0.1cm}
\begin{algorithm}[tbp]
  \caption{PO-based MO-PPO algorithm, AVUS}
  \label{action-MOPPO}
  \begin{algorithmic}[htbp]
  \REQUIRE ~~\\% Input
  PPO network structure, preference distribution $\mathcal{B}$, path $\Delta{\varpi}$ for coefficient $\varpi$.\\
  \ENSURE The optimal MO-PPO policy network.\\
  \STATE \textbf{Initialize:} Hyperparameters of PPO network, total epochs $\overline{U}$ in each update, minibatch size $M$, update frequency $\mathbf{\mathcal{U}}$ for MO-PPO algorithm.
  \FOR {iteration = 1, 2, $\cdots$}
  \STATE Sample a linear preference $\overline{\pmb{\omega}}$ from $\mathcal{B}$.
	\FOR {actor = 1, 2, $\cdots$, N}
	  \STATE Run policy $\pi_{\overline{\pmb{\theta}}}$ in environment for $T$ time steps.
	  \STATE Compute advantage estimates $\hat{A}_{1}, \cdots, \hat{A}_{\overline{T}}$ for every $\overline{T}$ updating time.
	\ENDFOR
    \STATE Optimize loss function $\mathcal{L}$ wrt $\overline{\pmb{\theta}}$, with $\overline{U}$ and $M \leq \mathbf{\mathcal{U}}$, according to equation \eqref{Loss function}.
    % $\mathcal{L}^{\mathrm{NCP/CLIP/KL}}(\overline{\pmb{\theta}}) = \varpi \mathcal{L}_{1}^{\mathrm{NCP/CLIP/KL}}(\overline{\pmb{\theta}}) + (1-\varpi) \mathcal{L}_{2}^{\mathrm{NCP/CLIP/KL}}(\overline{\pmb{\theta}})$.
    \STATE Update parameters $\overline{\pmb{\theta}} \leftarrow \overline{\pmb{\theta}}^{*}$.
    \STATE Increase $\varpi$ along the path $\Delta{\varpi}$.
  \ENDFOR
  \end{algorithmic}
\end{algorithm}
% \vspace{-0.3cm}

\noindent
The $\mathcal{L}^{\mathrm{Optimal}}_{1}(\overline{\pmb{\theta}}, \overline{\pmb{\omega}})$ is capable of ensuring that predicted action-value closing to any real expected total reward although it may not obtaining the optimal results. $\mathcal{L}^{\mathrm{Optimal}}_{2}(\overline{\pmb{\theta}}, \overline{\pmb{\omega}})$ is able to pull along the proper direction with better utility. Therefore, to obtain the optimal results, the final loss function can be expressed according to the homotopy optimization \cite{7}:
\par
\vspace{-0.1cm}
\noindent
\begin{align}\label{Loss function}
  \mathcal{L}^{\mathrm{Optimal}}(\overline{\pmb{\theta}}, \overline{\pmb{\omega}}) = \varpi \mathcal{L}^{\mathrm{Optimal}}_{1}(\overline{\pmb{\theta}}, \overline{\pmb{\omega}}) + (1-\varpi) \mathcal{L}^{\mathrm{Optimal}}_{2}(\overline{\pmb{\theta}}, \overline{\pmb{\omega}}),
\end{align}
\par
\vspace{-0.1cm}
\noindent
where $\varpi$ is a weight to trade off between $\mathcal{L}^{\mathrm{Optimal}}_{1}(\overline{\pmb{\theta}}, \overline{\pmb{\omega}})$ and $\mathcal{L}^{\mathrm{Optimal}}_{2}(\overline{\pmb{\theta}}, \overline{\pmb{\omega}})$. The value of $\varpi$ is increased from 0 to 1 with step 0.1. The pseudo-code of the AVUS-based MO-PPO algorithm is shown in \textbf{Algorithm~\ref{action-MOPPO}}. To sum up, the AVUS aims to train an agent to recover policies for approaching the entire PF. However, different preference $\overline{\pmb{\omega}}$ affects the total obtained rewards for coverage and capacity.

\vspace{-0.4cm}
\subsection{LFUS-based MO-PPO Algorithm}
\vspace{-0.1cm}
In this subsection, we consider the LFUS-based MO-PPO algorithm, where the multi-task learning (MTL) method is employed. Different from the AVUS, there are multiple gradient policies that need to be updated simultaneously. In the MTL-based MO-PPO problem, the empirical risk minimization formulation is generally followed:
\par
\vspace{-0.1cm}
\noindent
\begin{align}\label{empirical risk minimization}
  \min_{\overline{\pmb{\theta}}} \sum_{m=1}^{M} \varphi^m \hat{\mathcal{L}}^m(\overline{\pmb{\theta}}),
\end{align}
\par
\vspace{-0.1cm}
\noindent
where $\varphi^m$ and $\hat{\mathcal{L}}^m(\pmb{\overline{\theta}})$ denote the weights for $m$-th task and the empirical loss of $m$-th task. Consider two sets of solutions $\pmb{\overline{\theta}}_1$ and $\pmb{\overline{\theta}}_2$, if $\hat{\mathcal{L}}^{1}(\pmb{\overline{\theta}}_1) > \hat{\mathcal{L}}^{1}(\pmb{\overline{\theta}}_2)$ and $\hat{\mathcal{L}}^{2}(\pmb{\overline{\theta}}_1) < \hat{\mathcal{L}}^{2}(\pmb{\overline{\theta}}_2)$, it is obtained that the two tasks are mutually non-dominated, and therefore belong to the PF. In this case, the MTL problem can be formulated as MO optimization to explore the optimal results for conflicting objectives, where the vector-valued loss $\pmb{\mathcal{L}}$ are employed as follows:
\par
\vspace{-0.1cm}
\noindent
\begin{align}\label{vector loss}
  \min_{\pmb{\overline{\theta}}^{'}} \pmb{\mathcal{L}}(\pmb{\overline{\theta}}^{'}) = \min_{\pmb{\overline{\theta}}} [\hat{\mathcal{L}}^{1}(\pmb{\overline{\theta}}), \hat{\mathcal{L}}^{2}(\pmb{\overline{\theta}}), \cdots, \hat{\mathcal{L}}^{M}(\pmb{\overline{\theta}})]^T.
\end{align}
\par
\vspace{-0.1cm}
\noindent
Hence, the optimization of equation \eqref{vector loss} is to find PO solutions. Define $\overline{\mathcal{F}}=\{\pmb{\mathcal{L}}(\pmb{\overline{\theta}})\}, \pmb{\overline{\theta}} \in \pmb{\overline{\Theta}}$ as the approximate PF, where $\pmb{\overline{\theta}}$ and $\pmb{\overline{\Theta}}$ denote any one set of optimal parameters and all possible sets of optimal parameters. Here, we provide a theoretical analysis of the LFUS scheme below.

\subsubsection{Multiple Gradient Descent Algorithm (MGDA)}
To converge to the Pareto stationary (PS) solution problem, the MGDA \cite{8} is a proper method. According to the Karush-Kuhn-Tucker conditions, there exists $\nu_1,\nu_2,\cdots,\nu_M$ such that:
\begin{itemize}
  \item  $\nu_1,\nu_2,\cdots,\nu_M \geq 0$.
  \item  $\sum_{m=1}^{M}\nu_m = 1$ and $\sum_{m=1}^{M}\nu_m \nabla_{\overline{\pmb{\theta}}} \hat{\mathcal{L}}^m(\overline{\pmb{\theta}}) = 0$.
\end{itemize}
\par
Before handling the MGDA, the objectives may have values of the different scales, while MGDA is sensitive to the different ranges. Thus, the following gradient normalization method is invoked to alleviate the value range:
\vspace{-0.1cm}
\begin{align}\label{normalization}
  \nabla_{\overline{\pmb{\theta}}}\pmb{\mathcal{L}}(\overline{\pmb{\theta}}) = \frac{\nabla_{\overline{\pmb{\theta}}}\pmb{\mathcal{L}}(\overline{\pmb{\theta}})}{\pmb{\mathcal{L}}(\overline{\pmb{\theta}}^{'})},
\end{align}
\par
\vspace{-0.1cm}
\noindent
where $\overline{\pmb{\theta}}^{'}$ is the initial parameters of the model. Consequently, the range of the loss function has been limited to $[0, 1]$. 

\par
\vspace{-0.1cm}
\noindent
\begin{definition}\label{definition 1}
  A solution $\pmb{\overline{\theta}}_1$ dominates a solution $\pmb{\overline{\theta}}_2$ if for all objectives satisfying $\hat{\mathcal{L}}^{m}(\pmb{\overline{\theta}}_1) \leq \hat{\mathcal{L}}^{m}(\pmb{\overline{\theta}}_2)$, while exists at least one objective satisfying $\hat{\mathcal{L}}^{n}(\pmb{\overline{\theta}}_1) < \hat{\mathcal{L}}^{n}(\pmb{\overline{\theta}}_2)$, $\forall m,n \in \{1,2,\cdots,M\}$.
\end{definition}

\begin{definition}\label{definition 2}
  A solution $\pmb{\overline{\theta}}_1$ is PO solution while there is no any other solution $\pmb{\overline{\theta}}_2$ dominates $\pmb{\overline{\theta}}_1$.
\end{definition}

\begin{definition}\label{definition 3}
  All non-dominated solutions $\hat{\pmb{\overline{\theta}}}$ are Pareto set.
\end{definition}
\par
\vspace{-0.1cm}
\noindent
The solution that satisfies the conditions above is defined as a PS solution, while the PO solution is PS. Thus, the optimization problem can be defined as follows:
\par
\vspace{-0.1cm}
\noindent
\begin{align}\label{QCOP}
  \min_{\nu_1,\nu_2,\cdots,\nu_M} \Bigg\{ \bigg|\bigg|\sum_{m=1}^{M}\nu_m \nabla_{\overline{\pmb{\theta}}} \hat{\mathcal{L}}^m(\overline{\pmb{\theta}})\bigg|\bigg|^2_2 \Bigg| \sum_{m=1}^{M}\nu_m = 1, \nu_m \geq 0 \Bigg\},
\end{align}
\par
\vspace{-0.1cm}
\noindent
where $||\cdot||^2_2$ and $\nabla_{(\cdot)}$ denote the L2 norm and gradient descent (GD) operator. Define $\nabla_{\overline{\pmb{\theta}}^{'}} \mathcal{L}(\overline{\pmb{\theta}}^{'}) = \sum_{m=1}^{M}\nu_m \nabla_{\overline{\pmb{\theta}}} \hat{\mathcal{L}}^m(\overline{\pmb{\theta}})$, we have that: if $\nabla_{\overline{\pmb{\theta}}^{'}} \mathcal{L}(\overline{\pmb{\theta}}^{'}) = 0$, the solution is PS; otherwise, it isn't PS and $\nabla_{\overline{\pmb{\theta}}^{'}} \mathcal{L}(\overline{\pmb{\theta}}^{'})$ is the general GD vector. Since it has two objectives in problem \eqref{12}, the equation \eqref{QCOP} can be simplied as:
\par
\vspace{-0.1cm}
\noindent
\begin{align}\label{QCOP2}
  \min_{\nu \in [0,1]} ||\nu \nabla_{\overline{\pmb{\theta}}} \hat{\mathcal{L}}^1(\overline{\pmb{\theta}}) + (1-\nu) \nabla_{\overline{\pmb{\theta}}} \hat{\mathcal{L}}^2(\overline{\pmb{\theta}})||^2_2,
\end{align}
\par
\vspace{-0.1cm}
The optimization problem defined in \eqref{QCOP2} is equivalent to finding a minimum-norm point in the convex hull, which is a convex quadratic problem with linear constraints. Thus, an analytical solution to equation \eqref{QCOP2} can be expressed as:
\vspace{-0.1cm}
\begin{align}\label{QCOP2-solution}
  \nu = \bigg\{ \frac{[\nabla_{\overline{\pmb{\theta}}} \hat{\mathcal{L}}^2(\overline{\pmb{\theta}}) - \nabla_{\overline{\pmb{\theta}}} \hat{\mathcal{L}}^1(\overline{\pmb{\theta}})]^T\nabla_{\overline{\pmb{\theta}}} \hat{\mathcal{L}}^2(\overline{\pmb{\theta}})}{||\nabla_{\overline{\pmb{\theta}}} \hat{\mathcal{L}}^1(\overline{\pmb{\theta}}) - \nabla_{\overline{\pmb{\theta}}} \hat{\mathcal{L}}^2(\overline{\pmb{\theta}})||^2_2} \bigg\}_{[0,1]},
\end{align}
\par
\vspace{-0.1cm}
\noindent
where $\{\}_{[0,1]}$ represents clipping $\nu$ to $[0,1]$. Alternate optimization of GD vector and $\nu$ produces different $\nu$, which covers all PO solutions under constraints to approach the PF. According to the system model, it's suitable to select one PO solution as the optimal result.

\subsubsection{Loss Function}

% \vspace{-0.2cm}
\begin{algorithm}[htbp]
  \caption{PO-based MO-PPO algorithm, LFUS}
  \label{loss-MOPPO}
  \begin{algorithmic}[1]
  \REQUIRE ~~\\% Input
  PPO network structure.\\
  \ENSURE The optimal MO-PPO policy network.\\
  \STATE \textbf{Initialize:} Hyperparameters of PPO network, total epochs $\overline{U}$ in each update, minibatch size $M$, update frequency $\mathbf{\mathcal{U}}$ for MO-PPO algorithm.
  \FOR {iteration = 1, 2, $\cdots$}
    \FOR {objective = 1, 2, $\cdots$}
      \FOR {actor = 1, 2, $\cdots$, N}
        \STATE Run policy $\pi_{\overline{\theta}}$ in environment for $T$ time steps for each objective.
        \STATE Compute advantage estimates $\hat{A}_{1}, \cdots, \hat{A}_{\overline{T}}$ for each objective at every $\overline{T}$ updating time.
      \ENDFOR
    \ENDFOR
    \STATE Calculate loss function $\mathcal{L}$ wrt $\overline{\pmb{\theta}}$, with $\overline{U}$ and $M \leq \mathbf{\mathcal{U}}$, according to equation \eqref{optimal1}.
    \STATE Update $\overline{\pmb{\theta}}$ by min-norm solver.
  \ENDFOR
  \end{algorithmic}
\end{algorithm}
% \vspace{-0.3cm}

Our goal is to train one policy containing two sub-policies, where each objective has a specific loss function and shares all parameters. Thus, combing with the PPO algorithm, the loss function for the MO-PPO algorithm based on the NCP method, CLIP method, and KL Penalty method can be expressed as \eqref{loss function1} - \eqref{loss function3}, where $\hat{\mathbf{A}}_t$ is an advantage estimator, it can be expressed as \eqref{advantage}. Accordingly, at each update, the optimal method will be selected as follows:
\vspace{-0.1cm}
\begin{align}\label{optimal1}
  \mathcal{L}^{\mathrm{optimal}}(\overline{\pmb{\theta}}) = \max\{\mathcal{L}^{\mathrm{NCP}}(\overline{\pmb{\theta}}), \mathcal{L}^{\mathrm{CLIP}}(\overline{\pmb{\theta}}), \mathcal{L}^{\mathrm{KL}}(\overline{\pmb{\theta}})\}.
\end{align}
\par
\vspace{-0.1cm}
\noindent
The pseudo-code of the LFUS-based algorithm is shown in \textbf{Algorithm~\ref{loss-MOPPO}}. The proposed algorithm can be applied to other RIS-assisted scenarios, the configuration of networks is constructed according to the input and the training effect by the system model and formulated problem, while the AVUS and LFUS are still available to solve the problem once the formulated problem follows MO-MDP. For example, after providing the input and output of the problem formulated, the number of layers, neurons, and the optimizer can be adjusted for the training effect.

\par
\vspace{-0.1cm}
\noindent
\begin{remark}\label{remark 3}
  According to the theoretical analysis, the LFUS achieves a simpler structure than AVUS by only vectorizing the loss function. For AVUS, two policies are trained, since the action value is parallelly determined according to the preference according to \eqref{MO optimality filter-action}. For LFUS, only one policy is trained, since all the objectives share the same loss function in \eqref{vector loss}. Therefore, the LFUS should have a faster convergence speed than the AVUS.
\end{remark}
\par
\vspace{-0.1cm}
\noindent
\begin{figure*}[htbp]
  \normalsize 
\begin{align}\label{loss function1}
  \mathcal{L}^{\mathrm{NCP}}(\overline{\pmb{\theta}}) = \min_{\nu \in [0, 1]} \Big|\Big| \nu \mathbb{E}_{t}^{1}\Big\{\mathrm{min}\Big[\mathrm{\frac{\pi_{\overline{\pmb{\theta}}^{*}}(\mathbf{S}_t, \mathbf{A}_t)}{\pi_{\overline{\pmb{\theta}}}(\mathbf{S}_t, \mathbf{A}_t)}\hat{\mathbf{A}}_{t}^{\pi_{\overline{\pmb{\theta}}^{*}}}}\Big]\Big\} + (1-\nu) \mathbb{E}_{t}^{2}\Big\{\mathrm{min}\Big[\mathrm{\frac{\pi_{\overline{\pmb{\theta}}^{*}}(\mathbf{S}_t, \mathbf{A}_t)}{\pi_{\overline{\pmb{\theta}}}(\mathbf{S}_t, \mathbf{A}_t)}\hat{\mathbf{A}}_{t}^{\pi_{\overline{\pmb{\theta}}^{*}}}}\Big]\Big\}\Big|\Big|^{2}_{2},
\end{align}
\begin{align}\label{loss function2}
  &\mathcal{L}^{\mathrm{CLIP}}(\overline{\pmb{\theta}}) = \min_{\nu \in [0, 1]} \Big|\Big| \nu \mathbb{E}_{t}^{1}\Big\{\mathrm{min}\Big[\mathrm{\frac{\pi_{\overline{\pmb{\theta}}^{*}}(\mathbf{S}_t, \mathbf{A}_t)}{\pi_{\overline{\pmb{\theta}}}(\mathbf{S}_t, \mathbf{A}_t)}\hat{\mathbf{A}}_{t}^{\pi_{\overline{\pmb{\theta}}^{*}}},clip\Big(\frac{\pi_{\overline{\pmb{\theta}}^{*}}(S_t, A_t)}{\pi_{\overline{\pmb{\theta}}}(S_t, A_t)}, 1-\epsilon, \epsilon\Big)\hat{\mathbf{A}}_{t}^{\pi_{\overline{\pmb{\theta}}^{*}}}}\Big]\Big\} \nonumber \\
  &\hspace{6em} + (1-\nu) \mathbb{E}_{t}^{2}\Big\{\mathrm{min}\Big[\mathrm{\frac{\pi_{\overline{\pmb{\theta}}^{*}}(\mathbf{S}_t, \mathbf{A}_t)}{\pi_{\overline{\pmb{\theta}}}(\mathbf{S}_t, \mathbf{A}_t)}\hat{\mathbf{A}}_{t}^{\pi_{\overline{\pmb{\theta}}^{*}}},clip\Big(\frac{\pi_{\overline{\pmb{\theta}}^{*}}(S_t, A_t)}{\pi_{\overline{\pmb{\theta}}}(S_t, A_t)}, 1-\epsilon, \epsilon\Big)\hat{\mathbf{A}}_{t}^{\pi_{\overline{\pmb{\theta}}^{*}}}}\Big]\Big\}\Big|\Big|^{2}_{2},
\end{align}
\begin{align}\label{loss function3}
  &\mathcal{L}^{\mathrm{KL}}(\overline{\pmb{\theta}}) = \min_{\nu \in [0, 1]} \Big|\Big| \nu \mathbb{E}_{t}^{1}\Big\{\mathrm{min}\Big[\mathrm{\frac{\pi_{\overline{\pmb{\theta}}^{*}}(\mathbf{S}_t, \mathbf{A}_t)}{\pi_{\overline{\pmb{\theta}}}(\mathbf{S}_t, \mathbf{A}_t)}\hat{\mathbf{A}}_{t}^{\pi^{*}}, \tilde{\beta}KL(\pi_{\overline{\pmb{\theta}}^{*}}(\mathbf{S}_t), \pi_{\overline{\pmb{\theta}}}(\mathbf{S}_t))}\Big]\big\} \nonumber \\
  &\hspace{12em} + (1-\nu) \mathbb{E}_{t}^{2}\Big\{\mathrm{min}\Big[\mathrm{\frac{\pi_{\overline{\pmb{\theta}}^{*}}(\mathbf{S}_t, \mathbf{A}_t)}{\pi_{\overline{\pmb{\theta}}}(\mathbf{S}_t, \mathbf{A}_t)}\hat{\mathbf{A}}_{t}^{\pi^{*}},\tilde{\beta}KL(\pi_{\overline{\pmb{\theta}}^{*}}(\mathbf{S}_t), \pi_{\overline{\pmb{\theta}}}(\mathbf{S}_t))}\Big]\Big\}\Big|\Big|^{2}_{2},
\end{align}
\hrulefill \vspace*{0pt}
\end{figure*}

\begin{figure*}[htbp]
  \normalsize
\begin{align}\label{advantage}
  \hat{\mathbf{A}}_{t}^{\pi_{\overline{\pmb{\theta}}^{*}}} &= \sum_{t}^{\overline{T}}\mathbf{Q}_{\pi_{\overline{\pmb{\theta}}}}(\mathbf{S}_t, \mathbf{A}_t) - V_{\pi_{\overline{\pmb{\theta}}}}(\mathbf{S}_t) \nonumber \\
  &= \mathbf{R}_{t} + \gamma \mathbf{R}_{t+1} + \gamma^2 \mathbf{R}_{t+2} + \cdots + \gamma^{\overline{T}-t+1}\mathbf{R}_{\overline{T}-1} + \gamma^{\overline{T}-t}V_{\pi_{\overline{\pmb{\theta}}}}(\mathbf{S}_{\overline{T}}) - V_{\pi_{\overline{\pmb{\theta}}}}(\mathbf{S}_t).
\end{align}
\hrulefill \vspace*{0pt}
\end{figure*}

\vspace{-0.8cm}
\subsection{Empirical Complexity Analysis}
\vspace{-0.1cm}
As shown in Tab.~\ref{Ti}, we analyze the empirical complexity for the AVUS and LFUS, i.e., wall-clock time (time complexity) and memory utilization (space complexity). For the wall-clock time, AVUS spends 9.852s for 10 episodes and 16.42m for 1000 episodes, while it costs 8.934s for 10 episodes and 14.89m for 1000 episodes in LFUS. For memory utilization, LFUS consumes 108.35MB in total, which saves 7.33MB compared to AVUS. Therefore, the empirical complexity proves that LFUS can achieve less time complexity and space complexity than AVUS.
\vspace{-0.5cm}
\begin{table*}[!th]
  \caption{Resource footprint for AVUS and LFUS \label{Ti}}
  \centering
  \scalebox{0.85}{
  	\begin{tabular}{cccc}
  	\toprule
  	Policy & Time for 10 episodes & Time for 1000 episodes & Memory utilization in MB  \\
  	\midrule
  	AVUS & \textasciitilde9.852s & \textasciitilde16.42m & \textasciitilde115.68 \\
  	LFUS & \textasciitilde8.934s & \textasciitilde14.89m & \textasciitilde108.35 \\
  	\bottomrule
  	\end{tabular}
	}
\end{table*}

% \vspace{-0.1cm}
\section{Numerical Results}
% \vspace{-0.1cm}
In this section, we provide numerical results to evaluate the performance of proposed update strategies of MO-PPO algorithms. The simulation is fully performed on the CPU of the Dell Precision 7920 workstation. The configuration of the workstation is listed as follows: 1) CPU: Intel Xeon Bronze 3204 (8.25MB cache, 6 cores, 6 threads, up to 1.90GHz, 85W), 2) GPU: NVIDIA T400, 4GB GDDR6, 3) RAM: 8GB, 1x8GB, DDR4, 2933MHz, ECC, 4) ROM: 256GB, 2.5-inch, SATA, SSD, Class 20. Without loss of generality, a Poisson traffic model is employed to estimate the traffic flows or data sources for the proposed system model. In practice networks, there is a relationship in the traffic load between any adjacent time steps, where the traffic load at the current time step is determined by the previous time step. Based on this traffic model, the normalized coverage probability of observing $\overline{k}_i$ events and capacity probability of observing $\hat{k}_i$ events at sample point $s_i$ at time step 0 can be given by \cite{9}:
\vspace{-0.1cm}
\begin{align}\label{Poisson traffic model}
  w_{\mathrm{cov}, s_i}(0) = \overline{P}_{s_i}(\overline{k}_i) = \frac{e^{-\overline{\lambda}_i}\frac{-\overline{\lambda}_i^{\overline{k}_i}}{\overline{k}_i!}}{\sum_{i=1}^{N}e^{-\overline{\lambda}_i}\frac{-\overline{\lambda}_i^{\overline{k}_i}}{\overline{k}_i!}}, \hspace{1em} \\
  w_{\mathrm{cap}, s_i}(0) = \overline{P}_{s_i}(\hat{k}_i) = \frac{e^{-\hat{\lambda}_i}\frac{-\hat{\lambda}_i^{\hat{k}_i}}{\hat{k}_i!}}{\sum_{i=1}^{N}e^{-\hat{\lambda}_i}\frac{-\hat{\lambda}_i^{\hat{k}_i}}{\hat{k}_i!}},
\end{align}
\par
\vspace{-0.1cm}
\noindent
where $\overline{\lambda}_n$ and $\hat{\lambda}_n$ are the average number of events at each sample point $s_i$. The parameters of the MO-PPO network and communication network are given in Table.~\ref{Sim_PPO_network} and Table.~\ref{Sim_network}. The construction of both actor and critic networks are the same, the number of layers is three, which consists of the input layer, hidden layer, and output layer. The number of neurons in these three layers is $2N_s(2K+1)$, 64, $2N_s(\frac{P_{\mathrm{max}}+2}{z}+k-1)$, respectively. The optimizer is Adam while the parameters are randomly initialized. Additionally, there are two benchmarks conceived to evaluate the proposed update strategies:
\begin{itemize}
  \item \textbf{Without STAR-RIS (network performance)}: In this benchmark, the BSs serve the whole serving area without the assistance of the STAR-RISs.
  \item \textbf{Fixed weights (algorithm performance)}: In this benchmark, the weights of coverage and capacity are fixed as two cases: a) \textbf{BM1: weights 0.3 and 0.7}; and b) \textbf{BM2: weights 0.6 and 0.4}.
\end{itemize}

% \vspace{-0.5cm}
\begin{table*}[htbp]
  \captionsetup[]{font = {small}}
  \caption{Simulation parameters for MO-PPO algorithm \label{Sim_PPO_network}}
  \centering
  \begin{tabular}{ccc}
  \toprule
  Parameter & Description & Value  \\
  \midrule
  $\mathbf{\mathcal{E}}$ & The maximum number of episodes & 10000 \\
  $\mathbf{\mathcal{T}}$ & The maximum of time steps in each episode & 5000 \\ 
  $\mathbf{\mathcal{U}}$ & Update frequency for MO-PPO algorithm & 10 \\
  $\overline{U}$ & The number of epochs in each update & 10 \\
  $\overline{E}$ & Clipped parameter for MO-PPO algorithm & 0.2 \\
  $\eta$ & Discount factor & 0.99 \\
  $\psi_a$ & Learning rate for actor network & 0.0001 \\
  $\psi_c$ & Learning rate for critic network & 0.003 \\
  $\varpi$ & Initial coefficient for updating action-value strategy & 0.1 \\ 
  $\Delta{\varpi}$ & Step for the coefficient of updating action-value strategy & 0.001 \\
  \bottomrule
  \end{tabular}
  % \vspace{-0.4cm}  
\end{table*}

% \vspace{-1cm}
\subsection{Approximate PF by Different Proposed Strategies}
\vspace{-0.1cm}
As shown in Fig.~\ref{Front}, we provide the approximate PF under AVUS and LFUS. Among them, two approximate PFs are depicted are plotted at 3.5GHz and 26GHz signal frequencies using AVUS. Note that, the capacity and coverage are the cumulated results in a time period, where the optimized weights for coverage and capacity are dynamic in the proposed strategies. Compared to \textbf{BM1} and \textbf{BM2}, the coverage and capacity of the solutions on the two fronts both satisfy the PO definition, where at least one of them is better than the benchmarks. It is obtained that a dynamic combination for CCO in a time period is better than the fixed assignment of coverage and capacity. Moreover, the performance of different frequencies on STAR-RIS is an interesting question. When the system bandwidth is the same, mmWave is able to provide better capacity due to channel and frequency characteristics, while sub-6 GHz provides better coverage. Here, the channel model of the mmWave signal only considers the LOS component in \eqref{71} - \eqref{73}. According to the proposed strategy, we randomly select one result from approximate PF based on AVUS and LFUS for discussion.

\begin{table*}[tbp]
  \captionsetup[]{font = {small}}
  \caption{Simulation parameters for communication networks \label{Sim_network}} 
  \centering
  \begin{tabular}{ccc}  
  \toprule 
  Parameter & Description & Value  \\ 
  \midrule
  $\overline{\lambda}_{\mathrm{cov}}$ & Average number of events for coverage& 5 \\
  $\overline{\lambda}_{\mathrm{cap}}$ & Average number of events for capacity& 64 \\
  C & Path loss when d = 1m & -30dB \\
  $n^2$ & Noise power variance& 9$\times$10$^{\mathrm{-12}}$mW $\approx$ -140.46dBW\\
  $\mathrm{R}_{th}$ & Minimal RSRP for all sample points & 0.23mW $\approx$ -36.38dBW \\ 
  $P_{t,\mathrm{max}}$ & Maximum transmit power & 200mW = 23.01dBm \\
  $\alpha_{aR}$ & \tabincell{c}{Rician factor for channel from \\ $a$-th BS to $n_s$-th STAR-RISs} & 3dB \\
  $\alpha_{RP}$ & \tabincell{c}{Rician factor for channel from \\ $n_s$-th STAR-RISs to $s_i$-th sample point} & 3dB \\
  $\alpha_{aP}$ & \tabincell{c}{Rician factor for channel from \\ $a$-th BS to $s_i$-th sample point} & 3dB \\ 
  $\gamma_{a, n_s}$ & \tabincell{c}{Path loss factor for channel from \\ $a$-th BS to $n_s$-th STAR-RISs} & 3.5 \\ 
  $\gamma_{n_s, s_i}$ & \tabincell{c}{Path loss factor for channel from \\ $n_s$-th STAR-RISs to $s_i$-th sample point} & 2.8 \\
  $\gamma_{a, s_i}$ & \tabincell{c}{Path loss factor for channel from \\ $a$-th BS to $n_s$-th STAR-RISs} & 2.2 \\
  $z$ & Discrete step for amplitude of STAR-RISs & 7m \\
  $h_b$ & Height of BS & 7m \\
  $R_g$ & Length of each grid & 1m \\
  \bottomrule
  \end{tabular} 
  % \vspace{-0.4cm}  
\end{table*}

\vspace{-0.4cm}
\subsection{Convergence of MO-PPO Algorithm with Proposed Strategies}
\begin{figure}[htbp]
  % \vspace{-0.8cm}
  \centering
  \begin{minipage}[t]{0.45\textwidth}
  \centering
  \includegraphics[height = 2.35in, width = 3.2in]{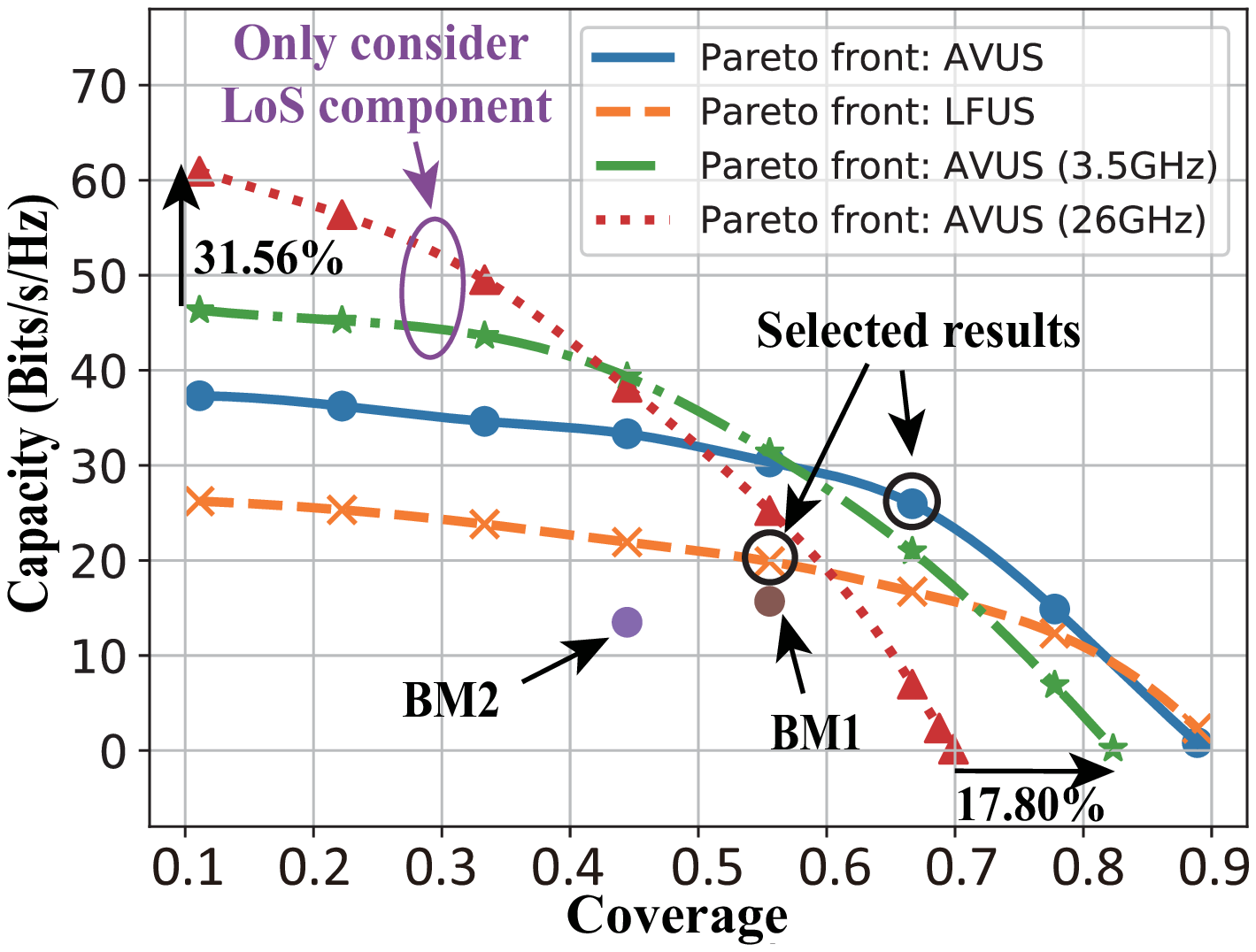}
  \caption{Approximate PF with different strategies, $N_s = 3$, $N = 9$, $K = 8 \times 10^2$, $I_{h_{n_s}}$ = 1.}  
  \label{Front}
  \end{minipage}\hspace{0.5cm}
  \begin{minipage}[t]{0.45\textwidth}
  \centering
  \includegraphics[height = 2.2in, width = 3.2in]{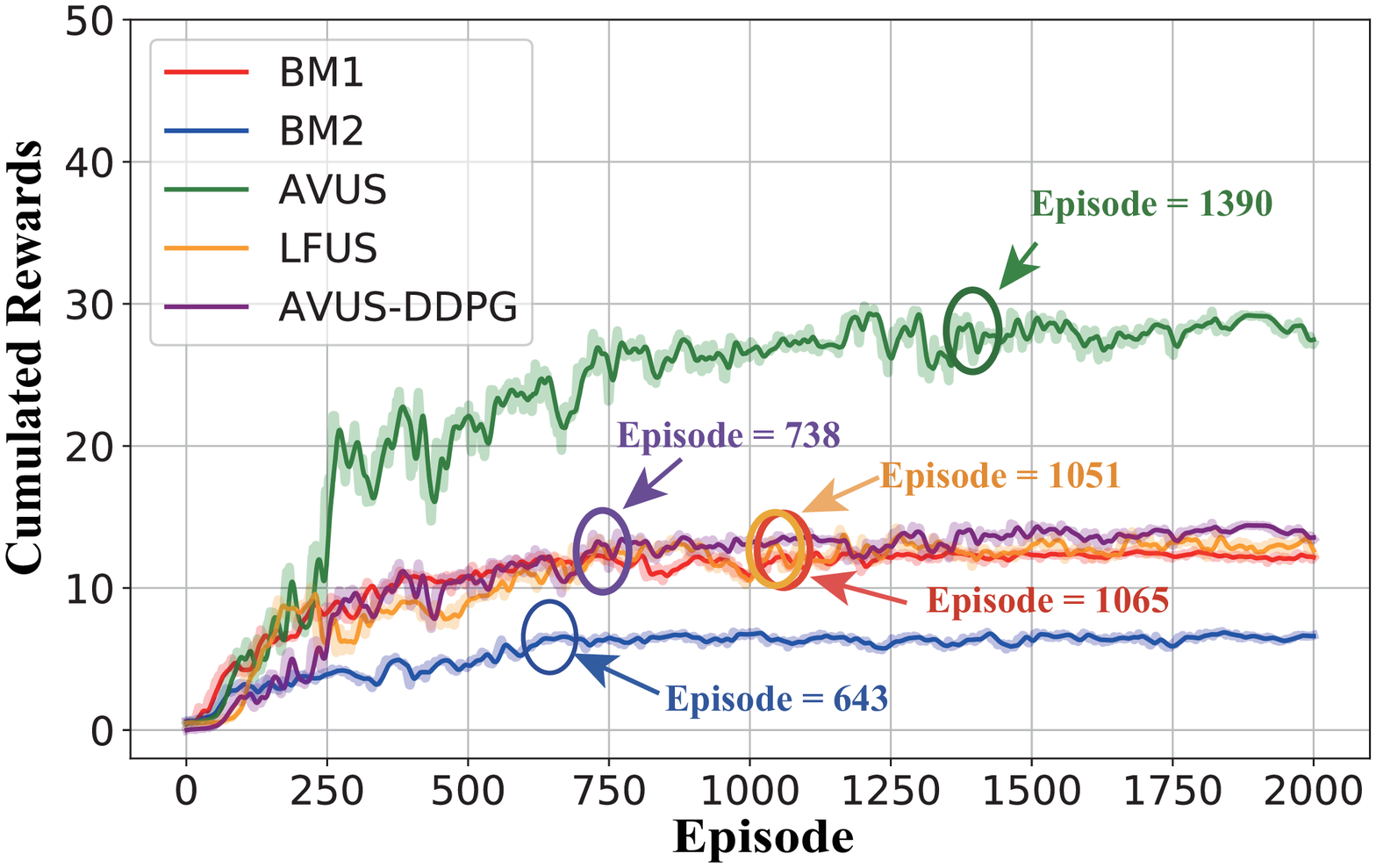}
  \caption{Learning curves (average convergence for ten times repeated training) for the MO-PPO algorithm with fixed weights, AVUS, and LFUS, initial $P_t$ = 2.1mW, $N_s = 3$, $N = 9$, $K = 8 \times 10^2$, $I_{h_{n_s}}$ = 1.}
  \label{learning_curve}
  \end{minipage}
\end{figure}

\vspace{-0.1cm}
In Fig.~\ref{learning_curve}, the convergence of the MO-PPO algorithm under proposed update strategies is demonstrated. Note that, to evaluate the performance of proposed algorithms, the learning curves are obtained by ten times repeated training. It can be observed from Fig.~\ref{learning_curve} that proposed strategies and benchmarks are capable of achieving convergence. Among them, the AVUS converges the slowest, but its cumulative reward is the largest, while the LFUS has a comparable convergence speed, but the cumulative reward is slightly higher. Compared to the benchmarks, both proposed algorithms are able to achieve better performance than the benchmarks in cumulated rewards or convergence speed. Furthermore, compared to other RL algorithms, e.g., deep deterministic policy gradient (DDPG), the DDPG has the fastest convergence speed and achieves the same performance as LFUS. This result proves the correctness of \textbf{Remark \ref{remark 3}} from practice.

\vspace{-0.3cm}
\subsection{Optimal Coverage and Capacity with Proposed Strategies}
\vspace{-0.1cm}
In this subsection, we will discuss the impact of the number of sample grids, the number of STAR-RISs, the number of elements in STAR-RISs, and the size of STAR-RISs on the selected optimal coverage and capacity.

% \vspace{-0.8cm}
\subsubsection{Impact of the Number of Sample Grids}
\begin{figure}[htbp]
  % \vspace{-0.8cm}
  % \setlength{\belowcaptionskip}{-0.9cm}
  \centering
  \subfigure[The optimized coverage under different numbers of sample grids of the serving area.]
  {
  \begin{minipage}[t]{0.45\textwidth}
  \centering
  \includegraphics[height=2.2in, width=3.2in]{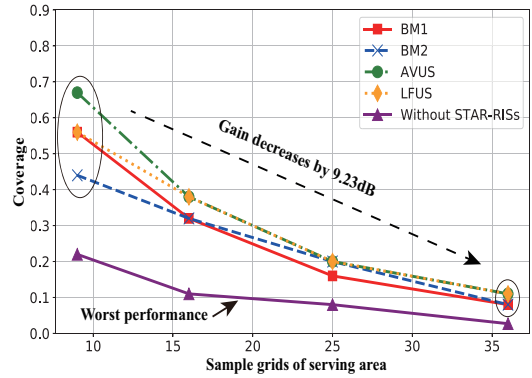}
  \label{optimized_cov_size}
  \end{minipage}
  }\hspace{0.75cm}
  \subfigure[The optimized capacity under different numbers of sample grids of the serving area.]
  {
  \begin{minipage}[t]{0.45\textwidth}
  \centering
  \includegraphics[height=2.2in, width=3.2in]{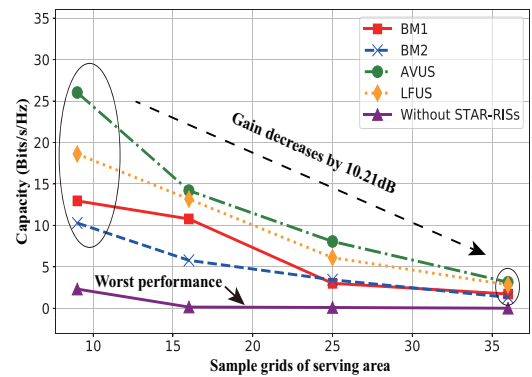}
  \label{optimized_cap_size}
  \end{minipage}
  }
  \caption{The optimized coverage and capacity for the MO-PPO algorithm with fixed weights, AVUS, and LFUS with sample grids $N$ of the serving area, $N_s = 3$, $K = 8 \times 10^2$, $I_{h_{n_s}}$ = 1.}
  \label{optimized_size}
\end{figure}

Fig.~\ref{optimized_size} characterizes the optimized coverage and capacity versus different total grids. In Fig.~\ref{optimized_cov_size}, it is observed that the coverage and capacity gains of all cases present decreasing trend with the upgrading of total grides. Specifically, the maximum decreasing gain of coverage among the proposed algorithms and fixed weight-based solutions is 9.23dB, while that capacity can achieve 10.21 dB. The reasons for these results are that compared with other sampling points, the fast fading channel characteristics of sampling points from BS and STAR-RISs make the received RSRP by far grids unable to reach $\mathrm{R}_{th}$. As a result, both coverage and capacity of the four cases present a downward trend. Additionally, compared to the "Without STAR-RISs" case, the proposed strategies and benchmarks show better performance. This is because the STAR-RISs proactively transmit and reflect the signal to the farther grid with less consumption. To sum up, in the case of only changing the total number of sampling points, the coverage and capacity changes are positively correlated with the total grid changes. Moreover, the proposed update strategies outperform the benchmarks, while the performance of the AVUS is better than the LFUS.

% \vspace{-0.4cm}

% \vspace{-0.4cm}
\begin{figure}[htbp]
  \centering
  \subfigure[The optimized coverage under different number of STAR-RISs.]
  {
  \begin{minipage}[t]{0.45\textwidth}
  \centering
  \includegraphics[height=2.2in, width=3.2in]{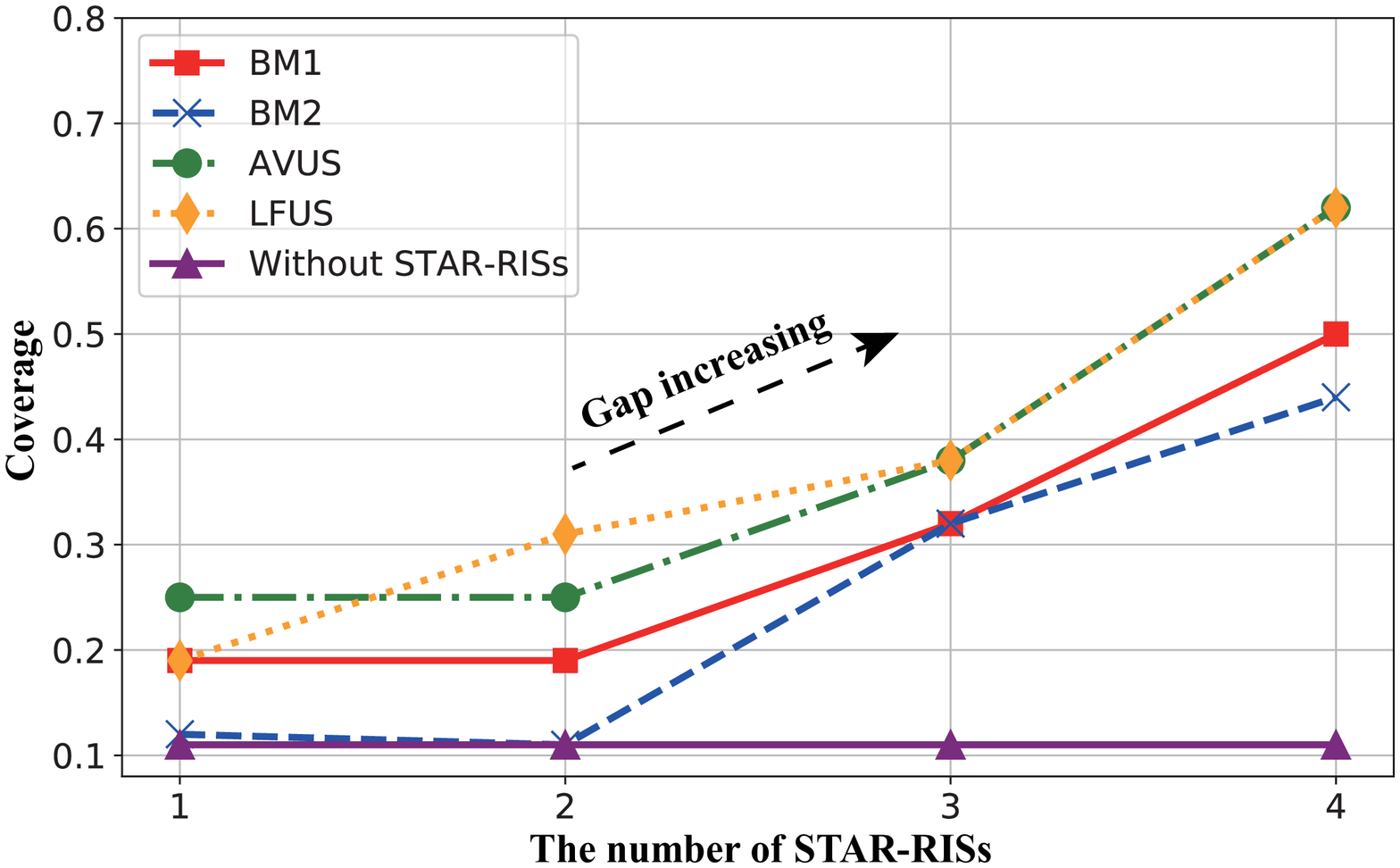}
  \label{optimized_cov_num}
  \end{minipage}
  }\hspace{0.75cm}
  \subfigure[The optimized capacity under different number of STAR-RISs.]
  {
  \begin{minipage}[t]{0.45\textwidth}
  \centering
  \includegraphics[height=2.2in, width=3.2in]{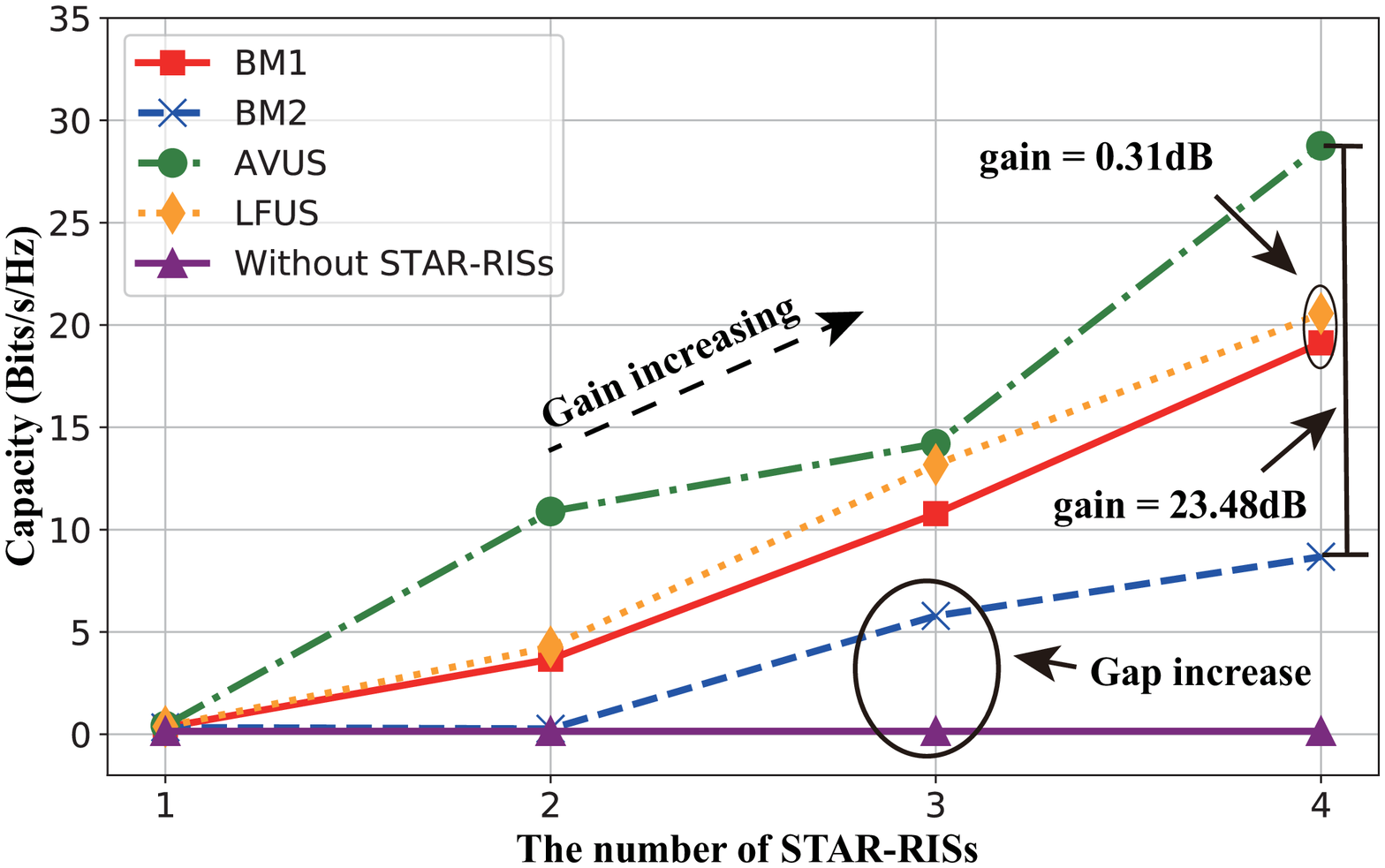}
  \label{optimized_cap_num}
  \end{minipage}
  }
  \caption{The optimized coverage and capacity for the MO-PPO algorithm with fixed weights, AVUS, and LFUS under different number $N_s$ of STAR-RISs, $N = 16$, $K = 8 \times 10^2$, $I_{h_{n_s}}$ = 1.}
  \label{optimized_num}
\end{figure}

\vspace{-0.3cm}
\subsubsection{Impact of the Number of STAR-RISs}
Fig.~\ref{optimized_num} depicts the optimized coverage and capacity versus the different numbers of STAR-RISs. As shown in Fig.~\ref{optimized_cov_num}, the coverage of all cases keeps growing steadily as the number of STAR-RISs increases. When the number of STAR-RISs $N_s$ reaches 4, the coverage of the \textbf{BM1} and \textbf{BM2} case can be promoted to over 0.4, and both proposed update strategies can arrive at over 0.6. This is because, with the increase in the number of STAR-RISs, STAR-RISs can help to compensate the received RSRP of some sample points to reach $\mathrm{R}_{th}$. For the capacity depicted in Fig.~\ref{optimized_cap_num}, the gain of capacity between AVUS and \textbf{BM2} case achieves 23.48dB, while the gain between LFUS and \textbf{BM1} only arrives at 0.31dB. This is because the STAR-RISs can compensate for the severe attenuation of channels from BSs to sample points, which indicates the effectiveness of STAR-RISs. Also, compared to the "Without STAR-RISs" case, the STAR-RISs are able to improve the coverage and capacity of the whole serving area. The gap between any multi-objective optimization solution (fixed weights or proposed strategies) and the "Without STAR-RISs" case keeps enlarging with the increase of the number of STAR-RISs. To sum up, it can be proved that the proposed update strategies also outperform the benchmarks for optimizing coverage and capacity. Since the STAR-RISs have presented their ability to improve spectrum utilization, the "Without STAR-RISs" case will not be discussed in the following subsections.

% \vspace{-0.8cm}
\begin{figure}[htbp]
  \centering
  \subfigure[The optimized coverage with different numbers of elements of STAR-RISs.]
  {
  \begin{minipage}[t]{0.45\textwidth}
  \centering
  \includegraphics[height=2.2in, width=3.2in]{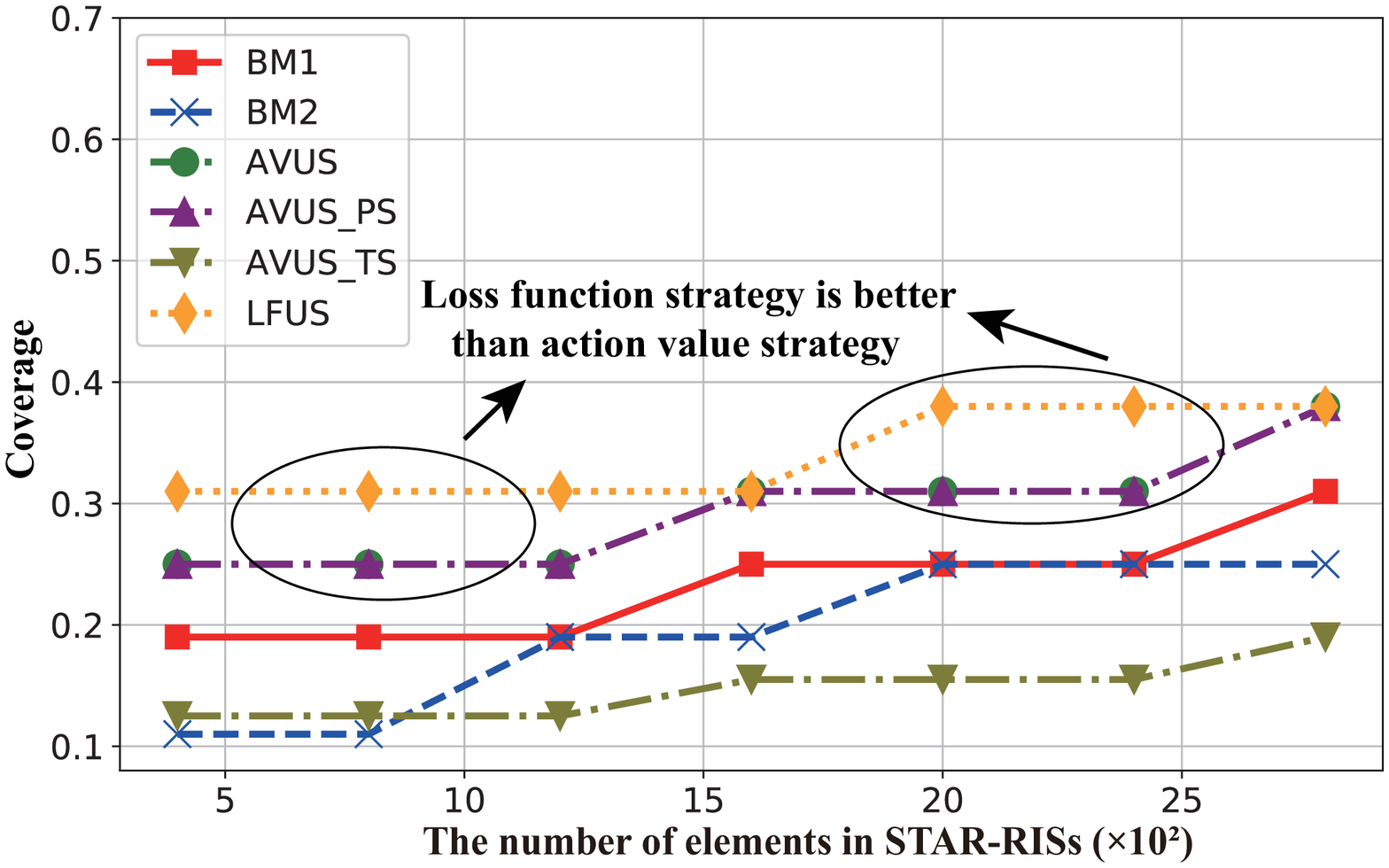}
  \label{optimized_cov_ele}
  \end{minipage}
  }\hspace{0.75cm}
  \subfigure[The optimized capacity with different numbers of elements of STAR-RISs.]
  {
  \begin{minipage}[t]{0.45\textwidth}
  \centering
  \includegraphics[height=2.2in, width=3.2in]{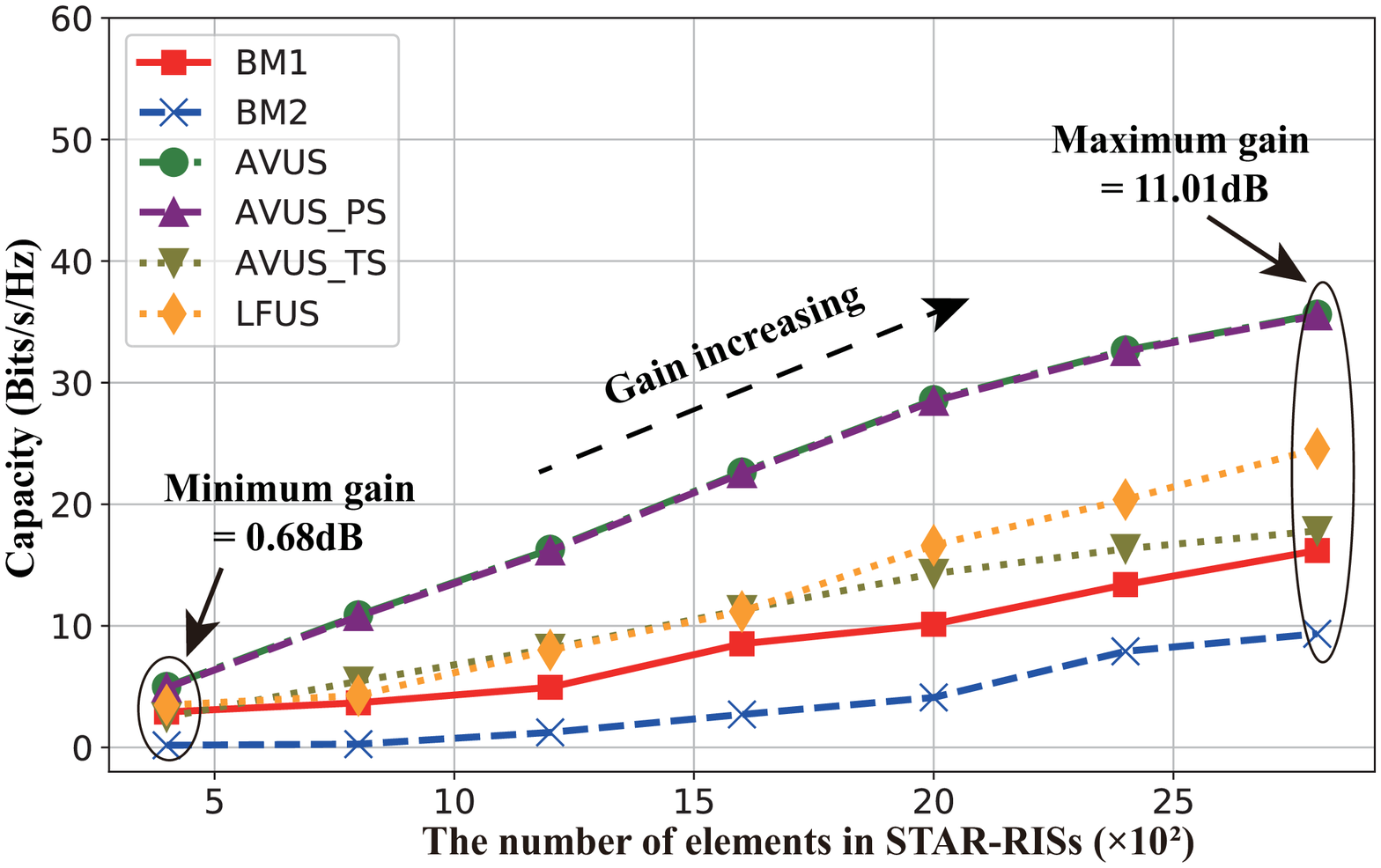}
  \label{optimized_cap_ele}
  \end{minipage}
  }
  \caption{The optimized coverage and capacity for the MO-PPO algorithm with fixed weights, AVUS, and LFUS with different numbers of elements $K$ of STAR-RISs, $N_s = 2$, $N = 16$, $I_{h_{n_s}}$ = 1.}
  \label{optimized_ele}
\end{figure}

% \vspace{-0.8cm}
\subsubsection{Impact of the Number of Element in STAR-RISs}
Fig.~\ref{optimized_ele} describes the optimized coverage and capacity versus the different number of elements in STAR-RISs. It can be observed that the coverage shows a slight change in Fig.~\ref{optimized_cov_ele}. The maximum gains among the optimized capacity of four cases in Fig.~\ref{optimized_cap_ele} are able to achieve 11.01dB when the number of elements in STAR-RISs increases to 36. It proves that the different number of elements in STAR-RISs bring a huge impact on optimizing capacity. This is because the role of each element is to transmit the BS signal to each sampling point while increasing the number of elements of STAR-RISs is adding multiple links to reduce loss. Compared with increasing the number of STAR-RISs, increasing the number of elements does not change the channel fast-fading characteristics of distant sample points. Also, in order to verify the effectiveness of the mode-splitting protocol in the system considered, we compare it with the PS and TS protocols. As shown in Fig.~\ref{optimized_ele}, the mode-splitting-based AVUS case outperformance the "AVUS-TS" case, while the gap between the mode-splitting-based AVUS optimization solution and the "AVUS-TS" case keeps enlarging with the increase in the number of STAR-RISs. This is because MS is able to make full use of the entire available communication time compared with TS. For the "AVUS-PS", there is no big difference between them. This is because the MS can be regarded as a special case PS. Moreover, for coverage, the LFUS outperforms the AVUS. It proves that when changing elements in STAR-RISs, the LFUS has a priority to be employed for only optimizing coverage. However, for both coverage and capacity optimization, it can be obtained that the AVUS is better than the LFUS. Also, the proposed update strategies both outperform the benchmarks.

% \vspace{-0.8cm}
\subsubsection{Impact of the Physical Size of STAR-RISs}

% \vspace{-0.4cm}
\begin{figure*}[htbp]
  % \setlength{\abovecaptionskip}{-0.1cm}
  % \setlength{\belowcaptionskip}{-0.8cm}
  % \centering
  \begin{minipage}[t]{0.45\textwidth}
  \subfigure[Case 1: The optimized coverage and capacity under different height of STAR-RISs, $\omega_{n_s} = 2$m.]
  {
  \includegraphics[height=2.2in, width=3.2in]{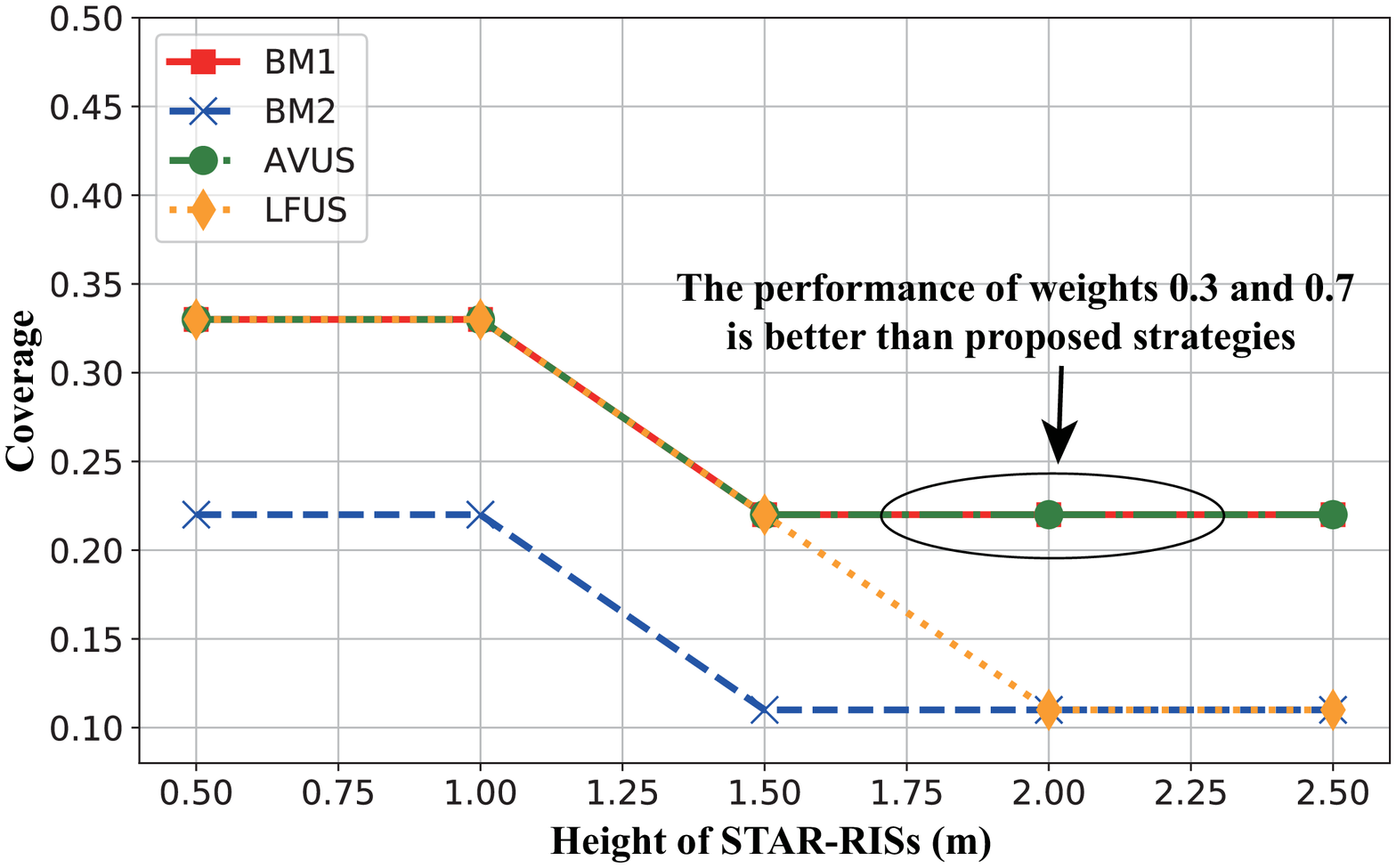}
  \hspace{3em}
  \includegraphics[height=2.2in, width=3.2in]{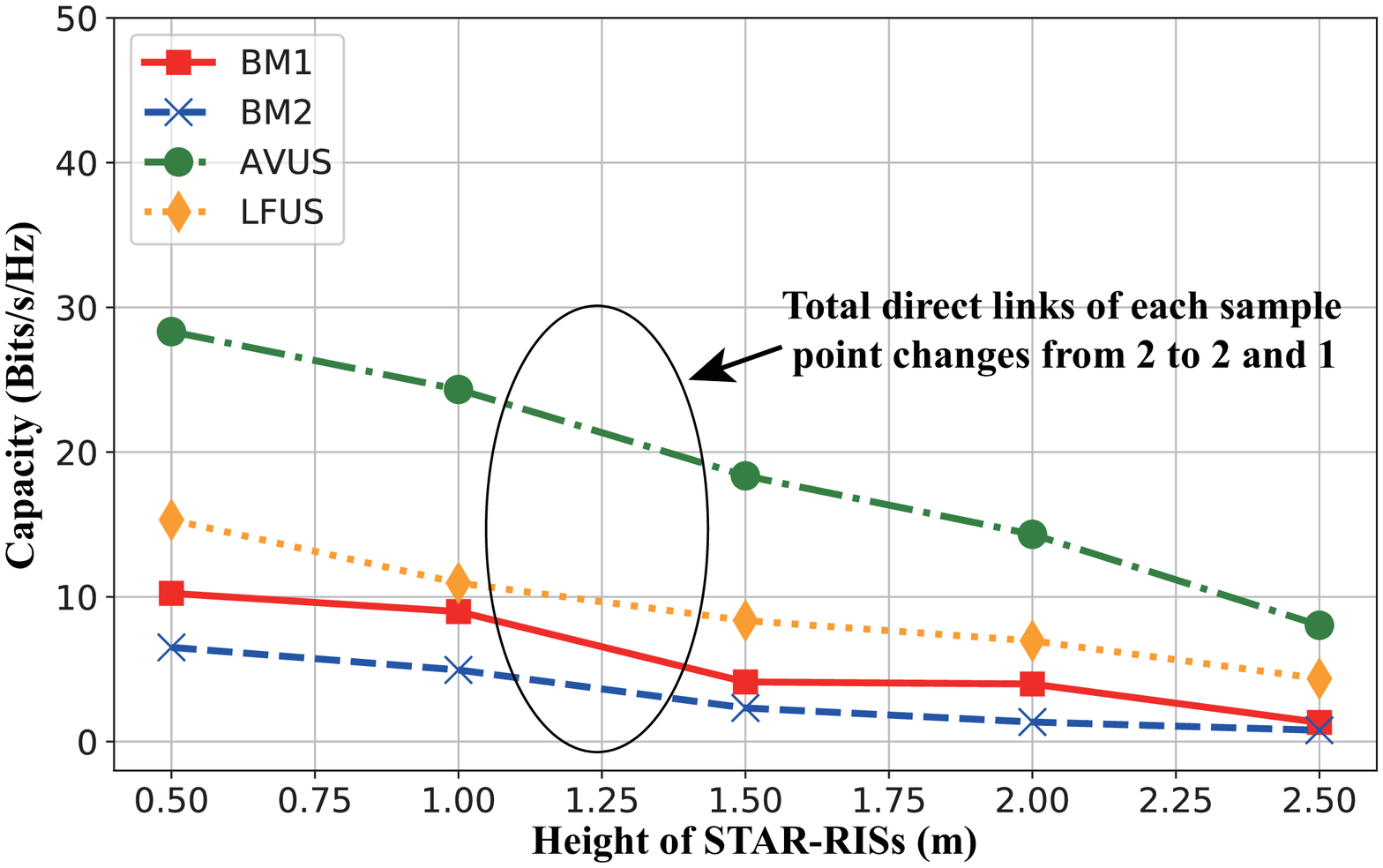}
  \label{wi2}
  }\end{minipage}
  \\
  \begin{minipage}[t]{0.45\textwidth}
  \subfigure[Case 2: The optimized coverage and capacity under different height of STAR-RISs, $\omega_{n_s} = 6$m.]
  {
  \includegraphics[height=2.2in, width=3.2in]{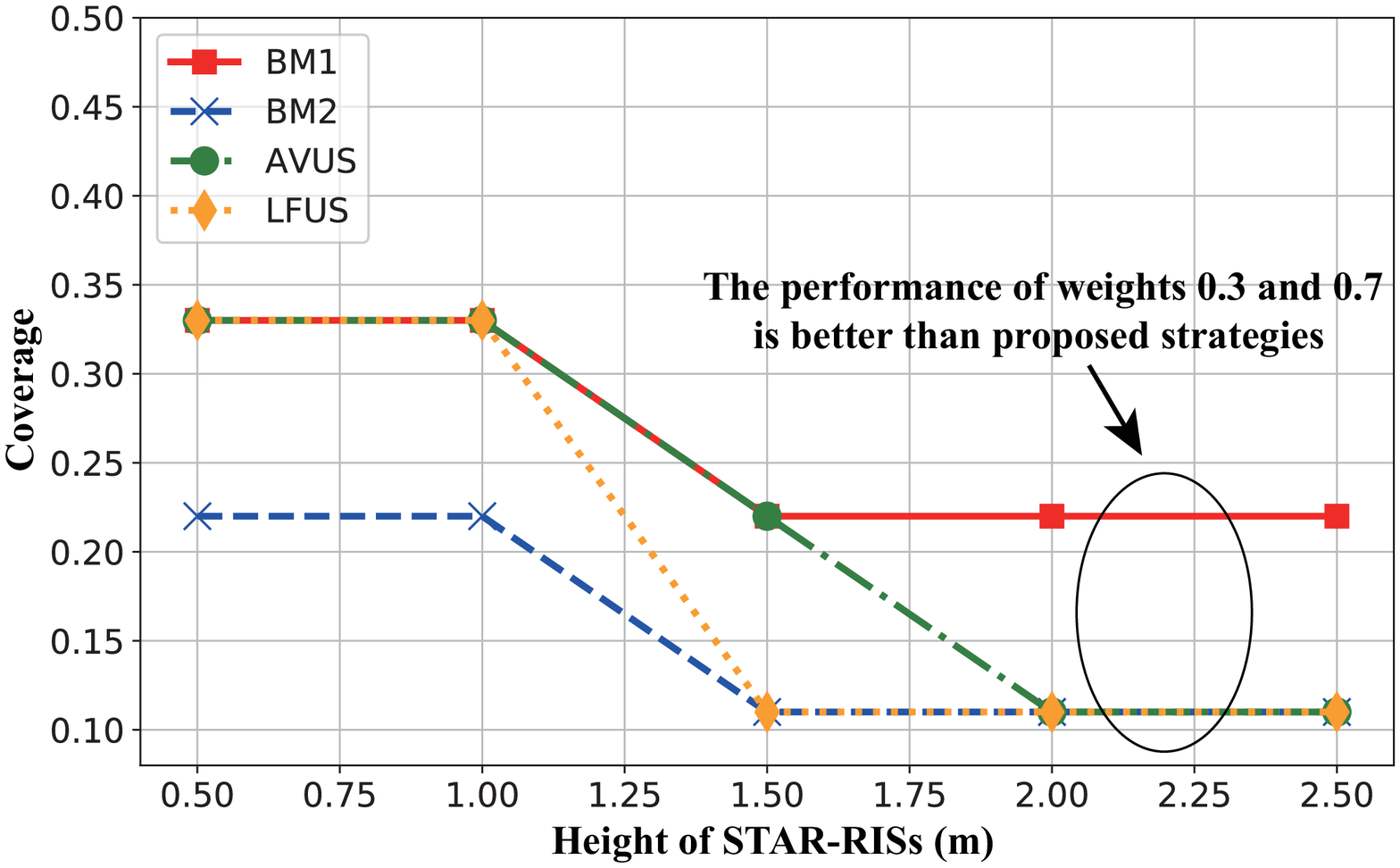}
  \hspace{3em}
  \includegraphics[height=2.2in, width=3.2in]{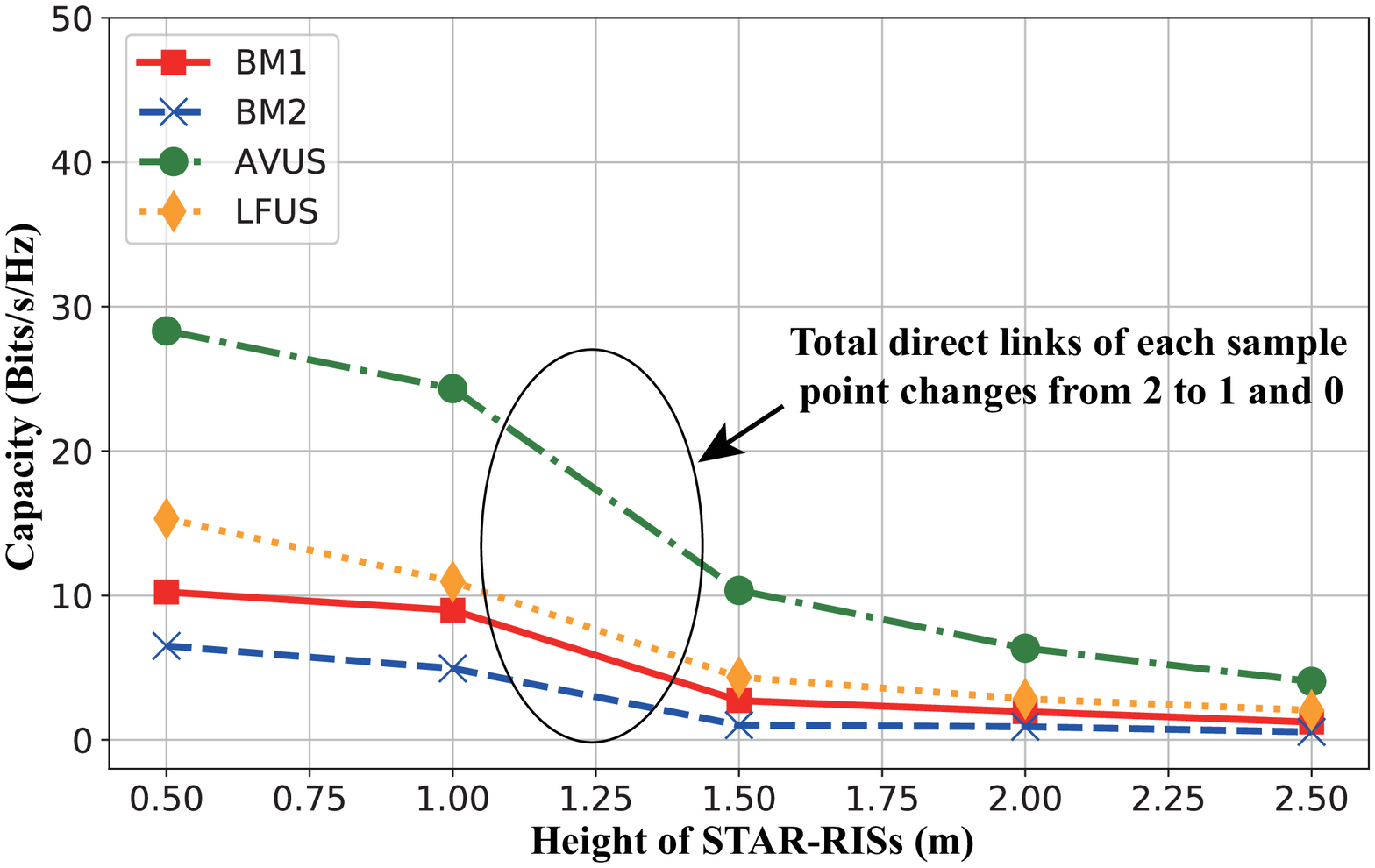}
  \label{wi6} 
  }\end{minipage}
  \\
  \begin{minipage}[t]{0.45\textwidth}
  \subfigure[Case 3: The optimized coverage and capacity under different width of STAR-RISs, $h_{n_s} = 2$m.]
  {
  \includegraphics[height=2.2in, width=3.2in]{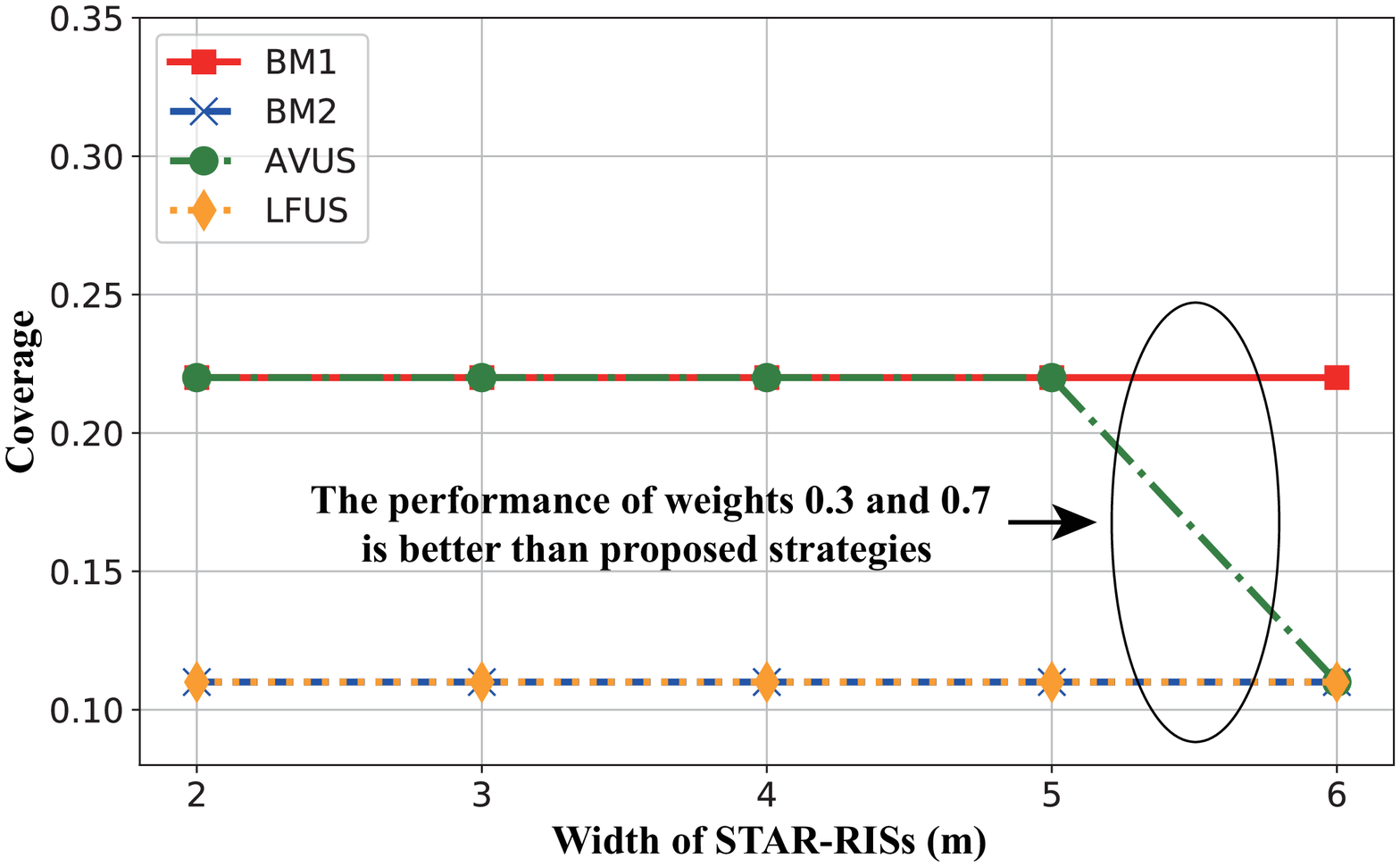}
  \hspace{3em}
  \includegraphics[height=2.2in, width=3.2in]{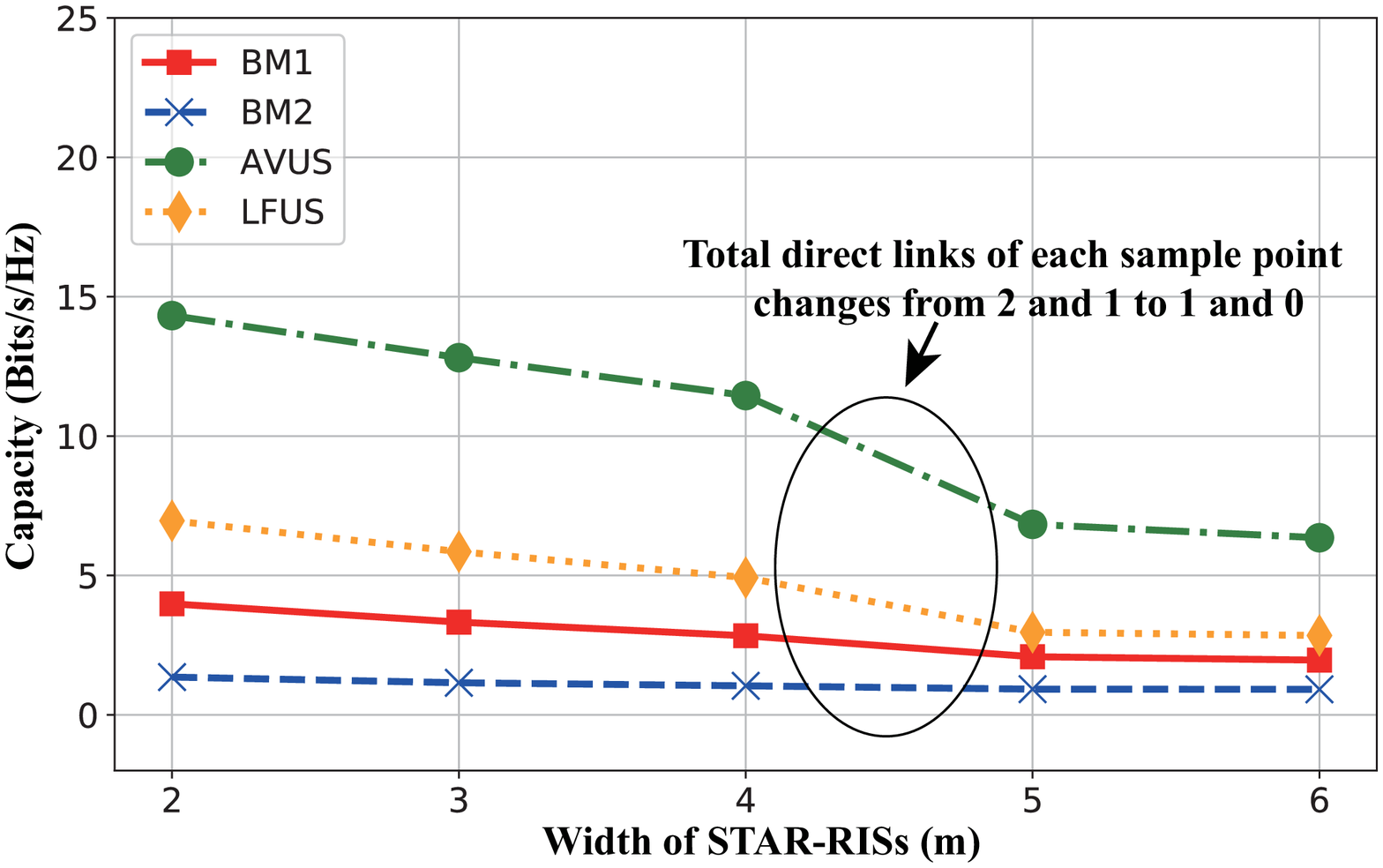}
  \label{he2}
  }\end{minipage}
  \caption{The optimized coverage and capacity for the MO-PPO algorithm with fixed weights, AVUS, and LFUS under different physical sizes of STAR-RISs, $N_s = 2$, $N = 16$.}
  \label{sizeRIS}
\end{figure*}

To evaluate the impact of the physical size of STAR-RISs on optimizing the coverage and capacity, the height $h_{n_s}$ and width $\omega_{n_s}$ of the STAR-RISs module are taken out for discussion. In this scenario, the number of STAR-RISs $N_s$, the number of total grids $N$, and the number elements in STAR-RISs $K$ are defined as: $N_s = 2$, $N = 16$. The number of elements K are increased linearly with the physical size of the STAR-RISs\footnote{The effective aperture of each STAR-RIS element keep the same as the area of each element won't be changed.}, and the $M_a = 6.25$ cm$^2$. According to the $h_b$ and $R_g$, the threshold of \eqref{STAR-RISs height threshold} and \eqref{STAR-RISs width threshold} can be calculated as 1m and 4m. Since $I_{h_{n_s}}$ = 1 has been discussed before, the other three scenarios are further considered as follows:
\begin{itemize}
  \item \textbf{Case 1}: Width of STAR-RISs are larger than the threshold, $I_{\omega_{n_s}}$ = 0
  \item \textbf{Case 2}: Width of STAR-RISs are smaller than the threshold, $I_{\omega_{n_s}}$ = 1
  \item \textbf{Case 3}: Height of STAR-RISs are smaller than the threshold, $I_{h_{n_s}}$ = 0
\end{itemize}
\par
Fig.~\ref{sizeRIS} demonstrate the optimized coverage and capacity for the MO-PPO algorithm with fixed weights, AVUS, and LFUS under the different physical sizes of STAR-RISs. Fig.~\ref{wi2} provides the changes of \textbf{Case 1}. In this case, there is at least one direct link between BS and any given sample point. When the height is also below the threshold, all sample points can have direct links with two BSs. Otherwise, one of the direct links among some sample points and BSs may be blocked. The coverage and capacity sharply fall down while the height of the STAR-RISs module passes over the threshold of 1m. This is because the number of direct links is a significant part to determine the strength of the received RSRP of each sample point. The upgrading number of direct links will increase the probability of reaching $R_{th}$ at each sampling point. For only considering capacity, the performance of proposed update strategies is better than benchmarks, while the AVUS outperforms the LFUS. For only considering coverage, the performance of proposed strategies cannot present better performance than \textbf{BM1}.

Fig.~\ref{wi6} provides the changes of \textbf{Case 2}. In this case, the number of direct links between BS and any given sample point can be 0, 1, and 2, which determines by the height of the STAR-RISs module. When the height is also below the threshold, all sample points can have direct links with two BSs. Otherwise, there is at most one direct link between sample points and BSs. The coverage and capacity dramatically decrease while the height of the STAR-RISs module passes over the threshold of 1m. This is because the locations of STAR-RISs determine that the direct links between the sample points and BSs are only 0 or 1. Different from the \textbf{Case 1}, the optimized coverage for the proposed update strategies is between benchmarks. It may indicate that the direct links play an important part in receiving RSRP, which needs to be further explored. Additionally, for only considering coverage, the performance of proposed strategies presents worse performance than \textbf{BM1}. But considering both coverage and capacity, the proposed update strategies are acceptable in \textbf{Case 2}.

Fig.~\ref{he2} provides the optimized coverage and capacity of \textbf{Case 3}. In this case, the height of the STAR-RISs module is fixed, which indicates that the direct links between sample points and BSs can be 0, 1, and 2. The number of direct links is 2 and 1, while the width of the STAR-RISs module is below the threshold. Otherwise, the number of direct links is 1 and 0. The capacity also shows a sharp falling down while the width goes over 4m. This is because the locations of STAR-RISs make the direct links between the sample points and BSs 0 or 1. Same with the \textbf{Case 2}, for only considering coverage, the performance of proposed strategies presents worse performance than \textbf{BM1}. Also, when considering both coverage and capacity, the proposed update strategies can be accepted.

\vspace{-0.4cm}
\section{Conclusion}
\vspace{-0.1cm}
In this paper, the coverage and capacity were modelled by considering the geographic property. Based on the model, we proposed a new framework for CCO in STAR-RIS-assisted wireless networks, by optimizing the transmit power, the reflection phase shift matrix, and the transmission phase shift matrix. In order to simultaneously optimize the coverage and capacity, an AVUS for the MO-PPO algorithm was investigated to solve the CCO problem, whose goal was to integrate action value for both coverage and capacity, which shared the same loss function. However, it had strict requirements on the computation resource thereby increasing the cost of the hardware. To handle this problem, another update strategy, i.e., the LFUS, was proposed to update the MO-PPO algorithm with an integrated loss function of coverage and capacity, whose goal was to consider the two-loss function for coverage and capacity. LFUS was able to dynamically assign the weights by a min-norm solver at each update for the MO-PPO algorithms. The numerical results proved that the investigated update strategies were able to provide more efficient solutions than the fixed-weight MOO algorithms. In addition, the coverage and capacity of wireless networks can be enhanced simultaneously with limited energy consumption since STAR-RISs had passive beamforming. In practice, multi-antenna BSs are usually deployed to improve the efficiency of the communication system by joint design active and passive beamforming, as well as considering the effect brought in practical imperfect CSI cases, which can be our future work on STAR-RIS-assisted networks.

\vspace{-0.3cm}
\begin{spacing}{1}
  
\end{spacing}

\end{sloppypar}
\end{document}